\documentclass[aps,prl,twocolumn,twoside,floatfix,superscriptaddress,preprintnumbers]{revtex4}
\usepackage{graphicx}

\newcommand{\bea}{\begin{eqnarray}}
\newcommand{\beq}{\begin{equation}}
\newcommand{\eea}{\end{eqnarray}}
\newcommand{\eeq}{\end{equation}}

\newcommand{\tgb}{ t_\beta }

\newcommand{\lsim}{\stackrel{<}{_\sim}}
\newcommand{\gsim}{\stackrel{>}{_\sim}}

\newcommand{\ba}{\begin{array}}
\newcommand{\ea}{\end{array}}

\newcommand{\Btaun}{{B_u \to \tau \nu}}

\begin{document}

\title{Waiting for $\mu\to e\gamma$ from the  MEG experiment}

\author{Junji~Hisano}
\affiliation{ICRR, University of Tokyo, Kashiwa 277-8582, Japan}
\affiliation{IPMU, University of Tokyo, Kashiwa, Chiba 277-8568, Japan}
\author{Minoru~Nagai}
\affiliation{Excellence Cluster Universe, Technische Universit\"at M\"unchen,
D-85748 Garching, Germany}
\author{P.~Paradisi}
\affiliation{Physik-Department, Technische Universit\"at M\"unchen,
D-85748 Garching, Germany}
\author{Yasuhiro Shimizu}
\affiliation{Tohoku University International Advanced Research and Education Organization, 
\\Institute for International Advanced Interdisciplinary Research, 
\\Sendai, Miyagi 980-8578, Japan}

\begin{abstract}
  The Standard Model (SM) predictions for the lepton flavor-violating (LFV) processes like
  $\mu\to e\gamma$ are well far from any realistic experimental resolution, thus,
  the appearance of $\mu\to e\gamma$ at the running MEG experiment would unambiguously
  point towards a New Physics (NP) signal. In this article, we discuss the phenomenological
  implications in case of observation/improved upper bound on $\mu\to e\gamma$ at the running
  MEG experiment for supersymmetric (SUSY) scenarios with a see-saw mechanism accounting
  for the neutrino masses. We outline the role of related observables to $\mu\to e\gamma$ in
  shedding light on the nature of the SUSY LFV sources providing useful tools {\it i)} to reconstruct
  some fundamental parameters of the neutrino physics and {\it ii)} to test whether an underlying SUSY
  Grand Unified Theory (GUT) is at work. The perspectives for the detection of LFV signals in
  $\tau$ decays are also discussed.
 
\end{abstract}

\maketitle

\section{Introduction}
\label{sec:introduction}

It is a well established fact that the Standard Model (SM) of the elementary particles represents a
very satisfactory model accounting for all the observed phenomena, both in (flavor-conserving)
electroweak (EW) physics at the LEP/SLC and also in low-energy flavor physics.

There are only few exceptions, as the anomalous magnetic moment $(g-2)$ of the muon and some
low-energy CP-violating observables measured at the $B$ factories that could indicate that the
SM might not be sufficient to describe them. Unfortunately, the hadronic uncertainties as well
as the limited experimental resolutions on flavor-changing neutral current (FCNC) data prevent
any conclusive evidence of New Physics (NP) effects in the quark sector.

In this respect, the FCNC phenomenology in the leptonic sector may be more promising. In fact,
neutrino physics has provided unambiguous indications about the non-conservation of the lepton
flavor, we therefore expect this phenomenon to occur also in the charged-lepton sector.

Interestingly, the charged LFV processes, such as $\mu\to e\gamma$, are severely suppressed
in the SM (with finite, but tiny neutrino masses) due to the GIM mechanism \cite{LFVnu}, hence, 
their observation would unambiguously point towards a NP signal arising from an underlying NP
theory operating at an energy scale not much above the TeV scale.
Similarly, also the leptonic EDMs represent very powerful and clean probes of NP effects
because of their high NP sensitivity and since their SM predictions are well far from any
realistic experimental resolution. (See Ref.~\cite{review} for a recent review about the
charged LFV processes and the EDMs.)

Despite of the great success of the SM, there is a general consensus that the SM has to be
regarded as an effective field theory, valid up to some still undetermined cut-off scale
$\Lambda$ above the EW scale. Theoretical arguments based on a natural solution of the
hierarchy problem suggest that $\Lambda$ should not exceed a few TeV, an energy scale that
would be explored at the upcoming LHC.

Supersymmetric (SUSY) extensions of the SM are broadly considered as the most natural and well
motivated scenario beyond the SM. Interestingly enough, the marriage of supersymmetry
and a see-saw mechanism accounting for the observed neutrino masses and mixing angles,
naturally leads to predictions for LFV processes as $\mu\to e\gamma$ well within
the experimental resolutions of the running MEG experiment at the PSI~\cite{meg}.

In this article, we discuss the implications of a potential evidence (or improved upper bound)
of ${\rm BR}(\mu\to e\gamma)$ at the expected sensitivities of MEG, namely at the level
of ${\rm BR}(\mu\to e\gamma)\gtrsim 10^{-13}$~\cite{meg}.

Assuming a supersymmetric framework, we exploit the correlations among
${\rm BR}(\mu\to e\gamma)$, the leptonic electric dipole moments
(EDMs) and the SUSY effects to the $(g-2)$ of the muon.  In case
$\mu\to e\gamma$ will be observed, we outline the complementary role
played by the leptonic EDMs and the P-odd asymmetry in $\mu^{+}\to
e^{+}\gamma$ to shed light on the nature of the LFV source. Finally,
the perspectives for the observation of LFV signals in $\tau$ decays
are also discussed.

\section{SUSY LFV and ${\rm BR}(\ell_i\to\ell_j\gamma)$}

Low-energy SUSY models generally contain new sources of flavor-violation in the soft-breaking parameters.
In particular, LFV effects relevant to charged leptons originate from any misalignment between fermion
and sfermion mass eigenstates. Once non-vanishing LFV entries in the slepton mass matrices are generated, irrespective to the underlying mechanism accounting for them, LFV rare decays like $\ell_i\to\ell_j\gamma$ are naturally induced by one-loop diagrams with the exchange of gauginos and sleptons.

In particular, the decay $\ell_i\to\ell_j\gamma$ is described by the dipole operator,
\begin{eqnarray}
{\cal L}_{\rm eff}= e\frac{m_i}{2}\overline{l}_i\sigma_{\mu\nu}
F^{\mu\nu}(A_L^{l_il_j}P_L+A_R^{l_il_j}P_R)l_j +h.c.,
\end{eqnarray}
and the decay rate is given by
\begin{eqnarray}
\frac{{\rm BR}(\ell_i\to\ell_j\gamma)}{{\rm BR}(\ell_i\to\ell_j\nu_i\bar{\nu}_j)}= \frac{48\pi^3\alpha_{em}}{G^2_F}\left(|A^{\ell_i\ell_j}_L|^2+|A^{\ell_i\ell_j}_R|^2\right)\,.
\end{eqnarray}
In the case where all the SUSY particles are degenerate, with a common mass $\tilde{m}$,
we find that
\begin{eqnarray}
  A^{\mu e}_L
  &=&
  \frac{\alpha_2}{4\pi} \frac{t_{\beta}}{\tilde{m}^{2}}
  \left[\frac{\delta^L_{\mu e}}{15}
    -\frac{\delta^L_{\mu\tau}\delta^L_{\tau e}}{40}
    -\frac{\alpha_Y}{\alpha_2}\frac{m_\tau}{m_\mu}
    \frac{\delta^R_{\mu \tau}\delta^L_{\tau e}}{30}
  \right]\,,
\\
  A^{\tau \ell}_L
  &=&
  \frac{\alpha_2}{4\pi} \frac{t_{\beta}}{\tilde{m}^{2}}
  \frac{\delta^L_{\tau \ell}}{15}\,,
\\
A^{\mu e}_R
&=&
-\frac{\alpha_Y}{4\pi}\frac{t_{\beta}}{\tilde{m}^{2}}
\left[\frac{\delta^R_{\mu e}}{60}
-\frac{\delta^R_{\mu \tau}\delta^R_{\tau e}}{60}
+\frac{m_\tau}{m_\mu}\frac{\delta^L_{\mu\tau}\delta^R_{\tau e}}{30}
\right]\,,
\\
A^{\tau \ell}_R
&=&
-\frac{\alpha_Y}{4\pi}\frac{t_{\beta}}{\tilde{m}^{2}}\frac{\delta^R_{\tau\ell}}{60}\,,
\end{eqnarray}
where $\tilde{m}$ is a typical SUSY mass running in the loop and $t_{\beta}=\tan\beta$
denotes the ratio of the two MSSM-Higgs vacuum expectation values (VEVs). Moreover, the
mass insertion (MI) parameters $\delta_{\ell_i\ell_j}^{L/R}$ for the left/right-handed
sleptons are defined as
\begin{eqnarray}
\delta^{L/R}_{\ell_i\ell_j} =
\frac{(m_{\tilde{L}/\tilde{R}}^2)_{\ell_i\ell_j}}{\tilde{m}^{2}}\,.
\label{MI}
\end{eqnarray}
Here, $(m_{\tilde{L}/\tilde{R}}^2)$ is the left/right-handed slepton mass matrix.

Besides $\ell_{i}\to\ell_{j}\gamma$, there are also other promising LFV channels, such as $\ell_i\to\ell_j\ell_k\ell_k$ and $\mu$-$e$ conversion in nuclei, that could be measured
with the upcoming experimental sensitivities. However, within SUSY models, these processes
are dominated by the dipole transition $\ell_{i}\to\ell_{j}\gamma^{*}$ leading to the
unambiguous prediction,
\begin{eqnarray}
{\rm BR}(\ell_{i}\rightarrow \ell_{j}\ell_{k}\ell_{k}) ~\sim~
\alpha_{\rm em} \times {\rm BR}(\ell_{i}\rightarrow \ell_{j}\gamma)\,,
\nonumber\\
{\rm CR}(\mu\rightarrow e~\mbox{in N}) ~\sim~
\alpha_{\rm em} \times {\rm BR}(\mu\rightarrow e\gamma)~.
\label{eq:dipole}
\end{eqnarray}
Thus, an experimental confirmation of the above relations would be crucial to prove
the dipole nature of the LFV transitions.

Additional contributions to LFV decays may arise from the Higgs sector through the effective LFV 
Yukawa interactions induced by non-holomorphic terms~\cite{bkl}. However, these effects become 
relevant only if $\tan\beta\sim\mathcal{O}(40-50)$ and if the Higgs masses are roughly one order
of magnitude lighter then the slepton masses~\cite{paradisiH}. The last condition never occurs
in our scenario hence Higgs mediated LFV effects are safely neglected in our analysis.
Sizable deviations from the expectations of Eq.~(\ref{eq:dipole}) may arise in case of large Higgs
mediated LFV effects~\cite{paradisiH}.

\section{The MSSM with right-handed neutrinos}
\label{sec:MSSM_seesaw}

As is well known, generic low-energy SUSY models with arbitrary soft-breaking terms would
induce unacceptably large flavor-violating effects. The unobserved departures from the SM
in quark FCNC transitions point toward the assumption of Minimal Flavor Violation~\cite{MFV}
or even flavor-universality in the SUSY-breaking mechanism. However, even under this assumption,
sizable flavor-mixing effects may be generated at the weak scale by the running of the
soft-breaking parameters from the (presumably high) scale of SUSY-breaking mediation~\cite{Hall:1985dx}.
In the leptonic sector, the relevance of such effects strongly depends on the assumptions
about the neutrino sector. If the light neutrino masses are obtained via a see-saw mechanism,
the induced flavor-mixing couplings relevant to LFV rates are naturally large~\cite{Borzumati:1986qx,Hisano:1995cp,Hisano:1998fj,Casas:2001sr,calibbi}.

Assuming a see-saw mechanism with three heavy right-handed neutrinos, the effective light-neutrino
mass matrix obtained integrating out the heavy fields is
\begin{equation}
\label{see-saw}
m_\nu = - Y_\nu \hat{M}^{-1} Y_\nu^T \langle H_u \rangle^2~,
\end{equation}
where $\hat{M}$ is the $3\times 3$ right-handed neutrino mass matrix (which breaks the
lepton number conservation), $Y_{\nu}$ are the $3\times 3$ Yukawa couplings between left-
and right-handed neutrinos (the potentially large sources of LFV), and $\langle H_u \rangle$
is the up-type Higgs VEV. Here, we take a basis where $\hat{M}$ is diagonal. Hereafter,
symbols with hat mean they are diagonal. Taking into account the renormalization-group
evolution (RGE), the slepton mass matrix $(m^2_{\tilde{L}})$ acquires LFV entries given by
\beq
(m^2_{\tilde{L}})_{ij} \simeq
- \frac{3m^2_0+A_0^2}{8\pi^2} (Y_{\nu})_{ik} (Y_{\nu}^\star)_{jk}
\ln \left(\frac{M_X}{M_{k}} \right)\,,
\label{Eq:dLL}
\eeq
with $i\neq j$ and $M_X$ denotes the scale of SUSY-breaking mediation while $m_0$ and $A_0$
stand for the universal SUSY breaking scalar mass and trilinear coupling at $M_X$, respectively.
Here, we assume $M_X$ is higher than the right-handed neutrino mass scale.
Starting from Eq.~(\ref{see-saw}), $Y_\nu$ can be written in the general form~\cite{Casas:2001sr}
$Y_\nu = U \sqrt{\hat{m}_\nu} R \sqrt{\hat{M}} / \langle H_u \rangle$ where $R$ is an arbitrary
complex orthogonal matrix while $U$ is the MNS matrix.

The MNS matrix contains three low-energy CP violating phases, the Dirac phase $\delta$
and two Majorana phases $\alpha$ and $\beta$. We use the standard parameterization~\cite{Maltoni:2004ei}:
\begin{eqnarray} \label{U_PMNS}
    U &\approx&\left(
    \begin{array}{ccc}
        c_{13} c_{12} & s_{12} c_{13} & s_{13}~ e^{-i \delta}  \\
       -s_{12} c_{23} & c_{23} c_{12} &  s_{23} c_{13} \\
        s_{23} s_{12} &-s_{23} c_{12} &  c_{23} c_{13}
    \end{array} \right)\nonumber\\
&\times&{\rm diag}(e^{i\alpha},e^{i\beta},1)\,,
\end{eqnarray}
where $c_{12}\!=\!\cos\theta_{12}$, $s_{12}\!=\!\sin\theta_{12}$, $c_{23}\!=\!\cos\theta_{23}$,
$s_{23}\!=\!\sin\theta_{23}$, $c_{13}\!=\!\cos\theta_{13}$ and $s_{13}\!=\!\sin\theta_{13}$ with
$s_{13}\lesssim 0.1$ (see Table~\ref{table:neutrinofit}).
 In Eq.~(\ref{U_PMNS}), we have systematically neglected all the subleading terms proportional
to $s_{13}$ but in the $U_{e3}$ matrix element where $s_{13}$ provides the leading contribution.
The Majorana phases $\alpha$ and $\beta$ are neglected in the following, except for the
cases in which they are relevant.

\begin{table} 
\begin{center}
\begin{tabular}{|c|}
\hline
Best fit values for light neutrinos\\
\hline
$\Delta m_{sol}^2 = (8.0 \pm 0.3) \times 10^{-5} ~\textrm{eV}^2$ \\
$|\Delta m_{atm}^2| = (1.9\mbox{--}3.0) \times 10^{-3} ~\textrm{eV}^2 \nonumber$ \\  
$\sin^2 2\theta_{12} = 0.86^{+0.03}_{-0.04}$\\
$\sin^2 2\theta_{23} >0.92$\\
$\sin^2 2\theta_{13} <0.19$\\
\hline
\end{tabular}
\caption{Best fit values for the light neutrino parameters~\cite{Amsler:2008zzb}.}
\label{table:neutrinofit}
\end{center}
\end{table}

A complete determination of $(m^2_{\tilde{L}})_{i\neq j}$ would require a complete knowledge of
the neutrino Yukawa matrix $Y_\nu$, which is not possible using only low-energy observables
from the neutrino sector~\footnote{ However, the marriage of the LHC data with potential LFV
signals from low-energy experiments, could provide additional tools to determine $Y_\nu$.
In particular, the off-diagonal terms in $(m^2_{\tilde{L}})_{ij}\sim (Y_\nu Y_\nu^{\dagger})_{ij}$
could be extracted from the branching ratios of LFV processes, as long as the relevant SUSY spectrum
is known. Then, the knowledge of both $(m^2_{\tilde{L}})_{ij}$ and the light neutrino mass matrix
might allow the determination of some parameters of the see-saw mechanism~\cite{seesaw}.}.
This is in contrast with the quark sector, where similar RGE contributions to the squark soft
masses are completely determined in terms of quark masses and CKM-matrix elements. As a result,
the predictions for FCNC effects in the lepton sector are usually affected by sizable uncertainties.

For future convenience, it is useful to write $(m_{{\tilde L}}^2)_{ij}$ in the following form
\begin{eqnarray}
(m_{{\tilde L}}^2)_{ij}\simeq - \frac{(3m_0^2+A_0^2)}{8\pi^2} U_{il}U^*_{jm} H_{lm}\,,
\end{eqnarray}
with $i\ne j$ and the Hermitian matrix $H_{lm}$ is defined as
\begin{eqnarray}
 H_{lm}=\frac{(m_{\nu_l}m_{\nu_m})^{1/2} \overline{M}_{k}}{{\langle H_u \rangle}^2}R_{lk} R^*_{mk}
\end{eqnarray}
with $\overline{M}_{k}=M_{k}\ln (M_X/M_{k})$.

We now discuss in detail the dependences of $(m_{{\tilde L}}^2)_{ij}$ on the parameters of
the neutrino sector. In spite of the many unknown parameters entering $(m_{{\tilde L}}^2)_{ij}$,
we note that the predictions for the correlations among LFV processes are affected by a much
smaller number of unknown parameters in some typical cases.
To see this point more explicitly, we consider the following ratio,
\beq
\frac{(m_{{\tilde L}}^2)_{e\mu}}{(m_{{\tilde L}}^2)_{\mu\tau}}= 
\frac{\sum_{ij} U_{e i} U^*_{\mu j} H_{ij}}{\sum_{ij} U_{\mu i} U^*_{\tau j} H_{ij}},
\label{d12d23}
\eeq
and we remind that the data from various neutrino experiments suggest that the MNS matrix contains
two large mixing angles, and only the $U_{e3}$ component can be small~\cite{Amsler:2008zzb}. Hence,
\begin{eqnarray}
 U_{e 3}U^*_{\mu i}\simeq U_{e 3}U^*_{\tau i}\simeq U_{e3}\ (i=1,2,3),
\end{eqnarray}
while all the remaining MNS matrix elements are ${\cal O}(1)$. If $H_{ij}$ does not have any
structure and all the components are comparable, Eq.~(\ref{d12d23}) implies that
\beq
 \label{Eq:LL_relation}
 (m_{{\tilde L}}^2)_{e \mu} \simeq (m_{{\tilde L}}^2)_{\mu \tau}\,.
\eeq
On the other hand, when only $H_{3i}~(i=1,2,3)$ provide the largest contributions, it turns out that
\beq
 \label{Eq:LL_relation2}
 \frac{(m_{{\tilde L}}^2)_{e \mu}}{(m_{{\tilde L}}^2)_{\mu\tau}}\simeq
 {\rm max}\{ U_{e3},\, H_{1j}/H_{3i},\, H_{2j}/H_{3i} \} \,.
\eeq
Finally, in both cases discussed above, we also find that
\beq
 (m_{{\tilde L}}^2)_{e\mu} \simeq (m_{{\tilde L}}^2)_{e\tau}\,.
\eeq
An experimental confirmation of these correlations would represent a powerful test for the
above scenarios, as well as a precious tool to shed light on some unknown neutrino parameters
of $H_{ij}$ and $U_{e3}$.

To make the above statements clear, we consider now the specific scenarios arising when $R=1$.
In this case, $H_{ij}$ contains only diagonal components and it takes the form
\begin{equation}
 H_{ij}= \frac{m_{\nu_i}\overline{M}_{i}}{{\langle H_u \rangle}^2} \delta_{ij}\,.
\end{equation}
The flavor mixing is controlled now only by three parameters, $H_{11}, H_{22}$ and $H_{33}$.
Even in this special case, the values of $H_{ii}$ are not uniquely defined as they still depend
on the unknown mass hierarchies for both light and heavy neutrinos.

Concerning the light neutrinos, we remind that in the hierarchical case one has
\beq
m_{\nu_2} - m_{\nu_1} \simeq m_{sol}~,  \qquad
m_{\nu_3} - m_{\nu_1} \simeq m_{atm}~,
\eeq
where we have assumed that $m_{\nu_1}\rightarrow 0$, while in the inverted hierarchy
case one has
\beq
m_{\nu_2} - m_{\nu_1} \simeq \frac{m^2_{sol}}{2 m_{atm}}~,
\qquad
m_{\nu_3} - m_{\nu_1} \simeq - m_{atm}~,
\eeq
where we have assumed the limit where $m_{\nu_3}\rightarrow 0$ (the notation is such that
$m_{atm}\!=\!\sqrt{|\Delta m^{2}_{atm}|}$ and $m_{sol}\!=\!\sqrt{\Delta m^{2}_{sol}}$).

Hence, in the following, we are lead with the following scenarios:
\begin{itemize}
 \item {\it normal hierarchy} for the light neutrinos and {\it hierarchical}
 right-handed neutrinos; $H_{ij}$ satisfies
\end{itemize}
\begin{eqnarray}
 H_{11} \ll H_{22} \ll H_{33}\,.
\label{neutrinos_nh}
\end{eqnarray}
Assuming that the off-diagonal elements of $(m_{{\tilde R}}^2)_{ij}$ are negligible,
Eq.~(\ref{neutrinos_nh}) implies that
\beq
\frac{{\rm BR}(\mu\to e\gamma)}{{\rm BR}(\tau\to\mu\gamma)}
\simeq
\frac{\left|\,s_{12}c_{12}(m_{sol}/m_{atm})(\overline{M}_{2}/\overline{M}_{3})+U_{e3}\,\right|^2}
{{\rm BR}(\tau\to\mu\nu_{\tau}\bar{\nu}_{\mu})/2}\,,
\label{mn_NH_MN_H}
\eeq
in agreement with the general expectation of Eq.~(\ref{Eq:LL_relation2}). Large values
for ${\rm BR}(\tau\to\mu\gamma)\lesssim 10^{-8}$, well within the reach of a Super $B$ factory~\cite{Akeroyd:2004mj}, are still possible provided $U_{e3}\lesssim 10^{-2}$
and $\overline{M}_{2}/\overline{M}_{3}\lesssim 0.1$ (see also Fig.~\ref{Fig:meg_tmg}).
\begin{itemize}
 \item {\it normal hierarchy} for the light neutrinos and
       {\it degenerate} right-handed neutrinos; $H_{ij}$ is such that
\end{itemize}
\begin{eqnarray}
 H_{11} \lsim H_{22} \lsim H_{33}\,,
\label{neutrinos_nd}
\end{eqnarray}
leading to the following result
\beq
\frac{{\rm BR}(\mu\to e\gamma)}{{\rm BR}(\tau\to\mu\gamma)}
\simeq
\frac{\left|\, s_{12}c_{12}\,(m_{sol}/m_{atm})+U_{e3}\,\right|^2}
{{\rm BR}(\tau\to\mu\nu_{\tau}\bar{\nu}_{\mu})/2}\,.
\label{mn_NH_MN_D}
\eeq
In contrast to the previous case, it is not possible now to reduce arbitrarily
${\rm BR}(\mu\to e\gamma)$ reducing the value of $U_{e3}$ because the contribution
from $H_{22}$ is not negligible. In fact, even setting $U_{e3}=0$, it turns out that
${\rm BR}(\tau\to\mu\gamma)\lesssim 10^{-10}\times({\rm BR}(\mu\to e\gamma)/10^{-11})$,
values well far from the expected experimental resolutions of a Super $B$ factory.
\begin{itemize}
 \item {\it inverted hierarchy} for the light neutrinos and {\it hierarchical}
       right-handed neutrinos; $H_{ij}$ is characterized by the following relation,
\end{itemize}
\begin{eqnarray}
 H_{11} \lsim H_{22}\,.
\label{neutrinos_ih}
\end{eqnarray}
In this scenario, the maximum allowed values for ${\rm BR}(\tau\to\mu\gamma)$ are obtained
assuming that $m_{\nu_3}\overline{M}_{3}\gg m_{\nu_2}\overline{M}_{2}$. In such a
case, it turns out that
\beq
\frac{{\rm BR}(\mu\to e\gamma)}{{\rm BR}(\tau\to\mu\gamma)}\simeq
\frac{\left|\,s_{12}c_{12}(m_{atm}/m_{\nu_3})(\overline{M}_{2}/\overline{M}_{3})+U_{e3}\,\right|^2}
{{\rm BR}(\tau\to\mu\nu_{\tau}\bar{\nu}_{\mu})/2}\,,
\label{mn_IH_MN_H}
\eeq
and large values for ${\rm BR}(\tau\to\mu\gamma)$ may be still allowed if $U_{e3}\lesssim 10^{-2}$
and depending on the unknown neutrino mass scale $m_{\nu_3}$ as well as the value of $\overline{M}_{2}/\overline{M}_{3}$.
\begin{itemize}
 \item {\it inverted hierarchy} for the light neutrinos and {\it degenerate} right-handed neutrinos;
the relation among the $H_{ij}$ elements is given by
\end{itemize}
\begin{eqnarray}
 H_{11}\simeq H_{22}\gsim H_{33}\,.
\label{neutrinos_id}
\end{eqnarray}
In this case, we find that
\beq
\frac{{\rm BR}(\mu\to e\gamma)}{{\rm BR}(\tau\to\mu\gamma)}
\simeq
\frac{\left|\,s_{12}c_{12}\,(m^{2}_{sol}/m^{2}_{atm})/2  - U_{e3}\,\right|^2}
{{\rm BR}(\tau\to\mu\nu_{\tau}\bar{\nu}_{\mu})/2}\,,
\label{mn_IH_MN_D}
\eeq
hence, large values for ${\rm BR}(\tau\to\mu\gamma)$ can be realized if $U_{e3}\lesssim 10^{-2}$.
Notice that, in Eq.~(\ref{mn_IH_MN_D}), the contributions from $H_{11}$ and $H_{22}$ to
${\rm BR}(\mu\to e\gamma)$ cancel each other to a very large extent. This implies that the
predictions for ${\rm BR}(\mu\to e\gamma)$ are highly sensitive to the degree of degeneracy
for the right-handed neutrino masses: even a modest mass splitting would imply a strong
enhancement for ${\rm BR}(\mu\to e\gamma)$.

In conclusion, large values for ${\rm BR}(\mu\to e\gamma)$ are possible in all the scenarios we
have discussed; in contrast, the attained values for ${\rm BR}(\tau\to\mu\gamma)$ are typically
very constrained by the current experimental bounds on ${\rm BR}(\mu\to e\gamma)$. The most
promising scenarios for $\tau\to\mu\gamma$ are those with {\it normal} or {\it inverted hierarchy}
for the light neutrinos and {\it hierarchical} right-handed neutrinos.
In any case, large values for ${\rm BR}(\tau\to\mu\gamma)$ would require $U_{e3}\lesssim 10^{-2}$.

Moreover, since $(m_{{\tilde L}}^2)_{e \mu} \simeq (m_{{\tilde L}}^2)_{e \tau}$, ${\rm BR}(\tau\to e\gamma)$
is always very suppressed in all the above scenarios once the current bound on ${\rm BR}(\mu\to e\gamma)$ is 
imposed.

One could wonder whether the picture we have outlined so far changes when we relax
the condition $R=1$. In this case, barring accidental cancellations, it seems difficult
to suppress simultaneously all the parameters of $H_{1i}$ and $H_{2i}$ $(i=1,2,3)$ while
keeping a large value for $H_{33}$. Thus, when $R\neq 1$, we end up with 
the generic prediction of Eq.~(\ref{Eq:LL_relation}),
irrespective to the details for the light and heavy neutrino masses.

\section{The SUSY $SU(5)$ model with right-handed neutrinos}
\label{sec:GUT}

More stable predictions for LFV effects in SUSY theories may be obtained embedding the SUSY model within
Grand Unified Theories (GUTs) where the see-saw mechanism can naturally arise (such as $SO(10)$).
In this case, the GUT symmetry allows us to obtain some hints about the unknown neutrino Yukawa matrix
$Y_{\nu}$. Moreover, in GUT scenarios there are additional flavor-violating contributions to slepton mass
terms stemming from the quark sector~\cite{Barbieri:1994pv,Barbieri:1995tw}. For instance, within $SU(5)$,
as both $Q$ and $e^c$ are hosted in the 10 representation, the CKM matrix mixing of the left-handed quarks
will give rise to off-diagonal entries in the running of the right-handed slepton soft masses $(m_{{\tilde R}}^2)_{ij}$ due to the interaction of the colored Higgs~\cite{Barbieri:1994pv,Barbieri:1995tw}.
These effects are completely independent from the structure of $Y_{\nu}$ and can be regarded as new
irreducible LFV contributions within SUSY GUTs. In particular, the expression for $(m_{{\tilde R}}^2)_{ij}$
within a SUSY $SU(5)$ model reads
\beq
(m_{{\tilde R}}^2)_{ij}\!=\!
-3\frac{(3m_0^2\!+\!A_0^2)}{8\pi^2}
(e^{i\hat{\phi}_d} V^T \hat{y}_u^2 V^* e^{-i\hat{\phi}_d})_{ij}
\ln\frac{M_X}{M_{\rm G}}\,,
\label{Eq:dRR}
\eeq
with $i\ne j$. Here, we have assumed the colored Higgs mass to be the GUT scale $M_{\rm G}$ and
$M_{\rm G}<M_X$. Moreover, $\hat{y}_u$ is the up-quark Yukawa coupling, $V$ is the CKM matrix and 
${\hat{\phi}}_{d}$ stands for additional physical CP-violating phases.

Within a pure SUSY $SU(5)$ model, where right-handed neutrinos are absent, only the flavor structures
$(m_{{\tilde R}}^2)_{ij}$ are at work and it turns out that
\beq
\frac{{\rm BR}(\mu\to e\gamma)}{{\rm BR}(\tau\to\mu\gamma)}\simeq
\frac{|V_{td}|^2}{{\rm BR}(\tau\to\mu\nu_{\tau}\bar{\nu}_{\mu})}\,,
\eeq
thus, ${\rm BR}(\tau\to\mu\gamma)\lesssim 2\times 10^{-8}\times({\rm BR}(\mu\to e\gamma)/10^{-11})$.
However, as we will show in the numerical analysis, all the processes ${\rm BR}(\ell_i\to\ell_j\gamma)$
turn out to be very suppressed in this scenario, except for few cases.
The main reasons for such a strong suppression are that {\it i)} the relevant sources of LFV, {\it i.e.}
$(m_{{\tilde R}}^2)_{ij}$, are CKM suppressed, {\it ii)} only $U(1)_Y$ interactions contribute to the LFV processes, {\it iii)} the total amplitude generating ${\rm BR}(\ell_i\to\ell_j\gamma)$ suffers from
strong cancellations in large regions of the parameter space \cite{Hisano:1996qq}.
As a result, large values for ${\rm BR}(\ell_i\to\ell_j\gamma)$ could be achieved only for light SUSY
particles and for moderate to large values of $\tan\beta$ and $A_0$; in this regime,
the indirect constraints, specially from the lower bound on the lightest Higgs boson mass and from
${\rm BR}(B\to X_s\gamma)$, become very strong and ${\rm BR}(\ell_i\to\ell_j\gamma)$ can hardly reach
the expected experimental resolutions.

The situation can drastically change if the SUSY $SU(5)$ model is enlarged to include right-handed
neutrinos ($SU(5)_{RN}$). In this case, besides the flavor structures $(m_{{\tilde R}}^2)_{ij}$ of Eq.~(\ref{Eq:dRR}), we also have the $(m_{{\tilde L}}^2)_{ij}$ MIs.

In the following, we assume a $SU(5)_{RN}$ model setting $R=1$ and assuming the scenario with
{\it normal hierarchy} for the light neutrinos and {\it hierarchical} right-handed neutrinos.
This leads to the following expression for $(m_{{\tilde L}}^2)_{ij}$
\begin{eqnarray}
(m_{{\tilde L}}^2)_{ij}\simeq -\frac{(3m_0^2+A_0^2)}{8\pi^2} U_{i3}U^*_{j3}
\frac{m_{\nu_3}\overline{M}_{3}}{{\langle H_u \rangle}^2}\,.
\end{eqnarray}
In the $SU(5)_{RN}$ model, the dominant contributions to $\mu\to e\gamma$ arise either from 
$\delta^L_{\mu e}$ (and $\delta^L_{\mu\tau}\delta^L_{\tau e}$) through the loop exchange of charginos/sneutrinos or from $\delta^L_{\mu\tau}\delta^R_{\tau e}$ through the loop exchange 
of a pure Bino; hence, ${\rm BR}(\mu\to e\gamma)$ can be written as ${\rm BR}(\mu\to e\gamma)\sim
a|\delta^{L}_{\mu e}|^2+b|\delta^{L}_{\mu\tau}\delta^{R}_{\tau e}|^2$~\cite{Hisano:1997tc}
with $a,b$ being functions of the SUSY parameters.
We note that, while $\delta^L_{\mu\tau}\delta^R_{\tau e}$ is $\sim U_{\mu3}V_{td}$ and thus
predictable in terms of known parameters, $\delta^L_{\mu e}$ depends on the unknown mixing
angle $U_{e3}$ as $\delta^L_{\mu e}\sim U_{e3}$. Concerning the correlation between
${\rm BR}(\mu\to e\gamma)$ and ${\rm BR}(\tau\to \mu\gamma)$, we can write
\beq
\frac{{\rm BR}(\mu\to e\gamma)}{{\rm BR}(\tau\to\mu\gamma)}
\simeq
\frac{\left(\, |U_{e3}|^{2},\, |\delta^{R}_{\tau e}|^2 \,\right)}
{{\rm BR}(\tau\to\mu\nu_{\tau}\bar{\nu}_{\mu})}\,,
\eeq
in case ${\rm BR}(\mu\to e\gamma)$ is dominated by $\delta^L_{\mu e}$ (if $U_{e3}\gtrsim 10^{-2}$)
or $\delta^{L}_{\mu\tau}\delta^{R}_{\tau e}$ (if $U_{e3}< 10^{-2}$), respectively.
In the latter case, we can expect large values for ${\rm BR}(\tau\to\mu\gamma)$ while taking
${\rm BR}(\mu\to e\gamma)$ easily under control as $|\delta^{R}_{\tau e}|\lesssim 10^{-3}$
(see Fig.~\ref{Fig:meg_tmg}).

In the numerical section, we will discuss about the predictions for LFV processes as arising both
in the pure SUSY $SU(5)$ model without right-handed neutrinos as well as in the $SU(5)_{RN}$ model.\\

In this article, we assume the minimal structure for the Yukawa couplings in
$SU(5)_{RN}$, for simplicity. However, more realistic models may introduce extra
contribution to flavor violation in the sfermion mass matrices.

It is known that the minimal SUSY $SU(5)$ GUT has two phenomenological problems:
{\it i) } the quark-lepton mass relations~\cite{Georgi:1979df} and {\it ii)} the proton decay
induced by the colored Higgs exchange~\cite{Sakai:1981pk}\cite{Goto:1998qg}.
The introduction of symmetries, like the Peccei-Quinn~\cite{pqextend} and the $U(1)_R$ 
symmetries~\cite{Hall:2001pg}, provides an economical solution to the problem {\it ii)},
as they suppress the baryon-number violating dimension-five operators induced by the
colored-Higgs exchange.
For the problem {\it i)}, the introduction of non-minimal Higgs/matter or higher-dimensional
operators with $SU(5)$-breaking Higgs has been proposed.
This proposal might be also a (partial) solution to the problem {\it ii)}, if the colored
Higgs coupling is accidentally suppressed~\cite{Bajc:2002bv}.
The new Higgs/matter or higher-dimensional interactions induce, in general, new sources
of flavor violation in the sfermion mass matrices. For example, sizable flavor mixings
in the left-handed slepton mass matrix might be generated even in the minimal SUSY SU(5)
GUT (without right-handed neutrinos) when higher-dimensional operators are introduced
to explain the quark-lepton mass relations~\cite{ArkaniHamed:1995fs}. Since these new
flavor violating interactions are assumed to be negligible in this article, our results
have to be regarded as conservative predictions of the $SU(5)_{RN}$ model, barring accidental
cancellations among different contributions.

%
\begin{table}[t]
\begin{center}
\begin{tabular}{||l|l||c|c||}
\hline
Observable & Present exp. bound \\
\hline
 ${\rm BR}(\mu\to e\gamma)$ & \qquad $1.2~ \times~ 10^{-11}$ \\
\hline
 ${\rm BR}(\mu\to eee)$ & \qquad $1.1~ \times~ 10^{-12}$\\
\hline
 ${\rm CR}(\mu \to e ~{\rm in ~Ti})$ & \qquad $4.3~ \times~ 10^{-12}$ \\
\hline
\hline
  ${\rm BR}(\tau \to \mu\,\gamma)$ & \qquad $4.5~ \times~ 10^{-8}$     \\
\hline
  ${\rm BR}(\tau \to e\,\gamma)$   & \qquad $1.1~ \times~ 10^{-7}$     \\
\hline
  ${\rm BR}(\tau \to \mu\,\mu\,\mu)$ & \qquad $1.9~ \times~ 10^{-7}$  \\
\hline
  ${\rm BR}(\tau \to \mu \eta)$     & \qquad $1.5~ \times~ 10^{-7}$    \\
\hline
\end{tabular}
\end{center}
\caption{\label{lfvtable}
Present experimental bounds on representative LFV decays of $\tau$ and $\mu$
leptons~\cite{Amsler:2008zzb}.}
\end{table}

\section{Muon $(g-2)$ vs ${\rm BR}(\ell_i\to\ell_j\gamma)$}
\label{sec:gm2_LFV}

Even if $\mu\to e\gamma$ will be observed, we could never access directly to the flavor-violating
parameters $\delta^{L,R}_{\ell_i\ell_j}$, since the branching ratio ${\rm BR}(\mu\to e\gamma)$
depends also on the SUSY particle masses and other parameters such as $\tan\beta$.
These latter parameters should be ultimately determined at the LHC/linear collider experiments
in the future. On the other hand, it is known that the SUSY effects to the muon $(g-2)$ are
well correlated with ${\rm Br}(\mu\to e\gamma)$ since they both are dipole transitions~\cite{Hisa1};
this is especially true in SUSY see-saw models~\cite{Hisa1} where the dominant effects to both
processes arise from the one-loop diagrams induced by chargino exchange.
Thus, normalizing ${\rm Br}(\mu\to e\gamma)$ to the SUSY effects to the muon $(g-2)$, we may get access
to the mass insertion parameters.

Most recent analyses of the muon $(g-2)$ converge towards a $3\sigma$ discrepancy in the $10^{-9}$ range~\cite{g_2_th}: $\Delta a_{\mu}\!=\!a_{\mu}^{\rm exp}\!-\!a_{\mu}^{\rm SM}\approx(3\pm 1)\times 10^{-9}$
where $a_{\mu}\!=\!(g-2)/2$. Despite substantial progress both on the experimental~\cite{g_2_exp}
and on the theoretical sides, the situation is not completely clear yet. However, the possibility
that the present discrepancy may arise from errors in the determination of the hadronic leading-order contribution to $\Delta a_{\mu}$ seems to be unlikely, as recently stressed in Ref.~\cite{passera_mh}.

The SUSY contribution to the muon $g-2$, $\Delta a^{\rm SUSY}_{\mu}$, in the limit of a degenerate SUSY spectrum reads
\beq
\Delta a^{\rm SUSY}_{\mu}\simeq
\frac{\alpha_2}{8\pi}\,\frac{5}{6}\frac{m^{2}_{\mu}}{\tilde{m}^{2}}\,t_{\beta}\,.
\eeq
For a natural choice of the SUSY parameters $t_{\beta}=10$ and $\tilde{m}=300\,{\rm GeV}$,
it turns out that $\Delta a^{\rm SUSY}_{\mu}\simeq 1.5 \times 10^{-9}$ and the current observed anomaly
can be easily explained.

Now, let us discuss the correlation between $\Delta a^{\rm SUSY}_{\mu}$ and the branching ratios of
$\ell_i\to\ell_j\gamma$.
Given our ignorance about the MI parameters $\delta^{L,R}_{\ell_i\ell_j}$, we will first perform a model-independent analysis, treating the MIs $\delta^{L,R}_{\ell_i\ell_j}$ as free parameters and
analysing their phenomenological impact separately. In particular, we will consider two cases for
$\mu\to e\gamma$, {\it i)} $\delta^L_{\mu e}$/ $\delta^L_{\mu\tau}\delta^L_{\tau e}$ dominance,
and {\it ii)} $\delta^L_{\mu\tau}\delta^R_{\tau e}$/ $\delta^R_{\mu\tau}\delta^L_{\tau e}$ dominance.

Assuming a degenerate SUSY spectrum, it is straightforward to find
\begin{eqnarray}
{\rm BR}(\mu\to e\gamma)&\approx&
2\times 10^{-12}
\left[\frac{\Delta a^{\rm SUSY}_{\mu}}{ 3 \times 10^{-9}}\right]^{2}
\bigg|\frac{\delta^{L}_{\mu e}}{10^{-4}}\bigg|^2\,,\nonumber\\
{\rm BR}(\mu\to e\gamma)&\approx&
3\times 10^{-13}
\left[\frac{\Delta a^{\rm SUSY}_{\mu}}{ 3 \times 10^{-9}}\right]^{2}
\bigg|\frac{\delta^{L}_{\mu\tau}\delta^{L}_{\tau e}}{10^{-4}}\bigg|^2\,,
\nonumber\\
{\rm BR}(\mu\to e\gamma)&\approx&
2\times 10^{-11}
\left[\frac{\Delta a^{\rm SUSY}_{\mu}}{ 3 \times 10^{-9}}\right]^{2}
\bigg|\frac{\delta^{L}_{\mu\tau}\delta^{R}_{\tau e}}{10^{-4}}\bigg|^2
\nonumber\\
&&+(L\leftrightarrow R),
\nonumber\\
{\rm BR}(\tau\to\ell\gamma)&\approx&
8\times 10^{-8}
\left[\frac{\Delta a^{\rm SUSY}_{\mu}}{ 3 \times 10^{-9}}\right]^{2}
\bigg|\frac{\delta^{L}_{\tau\ell}}{10^{-2}}\bigg|^2\,.
\end{eqnarray}
The main message from the above relations is that, as long as we intend to explain the muon $(g-2)$
anomaly within SUSY theories, the branching ratios for $\ell_i\to\ell_j\gamma$ are determined,
to some extent, once we specify the LFV sources. In the numerical section, we will address
this point in more details.

In Table~\ref{lfvtable2}, we report the bounds on the MIs $\delta_{l_i l_j}^{L}$ arising from the
current experimental bounds on ${\rm BR}(\ell_i\to\ell_j\gamma)$ imposing
$\Delta a^{\rm SUSY}_{\mu}=3\times 10^{-9}$, corresponding to the central value of the muon $(g-2)$ anomaly.
Moreover, in the last column of Table~\ref{lfvtable2} we also show the expectations for the MIs
$\delta^{L}_{l_i l_j}$ within the $SU(5)_{RN}$ scenario with hierarchical right-handed neutrinos for the
reference values ${M}^{\rm (r)}_{3}= 10^{13}\,{\rm GeV}$ and ${U}^{\rm (r)}_{e3}=0.1$. The predictions
for $\delta_{\mu e}^{L}$ and $\delta_{\tau e}^{L}$ scale with $U_{e3}$ as $(U_{e3}/{U}^{\rm (r)}_{e3})$
while all the $\delta_{l_i l_j}^{L}$'s scale with $M_{3}$ as $(M_{3}/{M}^{\rm (r)}_{3})$.
Table~\ref{lfvtable2} shows that, once we explain the muon $(g-2)$ anomaly through SUSY effects,
the current experimental resolutions on ${\rm BR}(\ell_i\to\ell_j\gamma)$ already set tight constraints
on the neutrino parameters $U_{e3}$ and $M_{3}$: if $U_{e3}$ is close to its experimental upper bound,
{\it i.e.} if $U_{e3}=0.1$, we are lead with ${M}_{3}\lesssim 10^{13}\,{\rm GeV}$.
\begin{table}[t]
\begin{center}
\begin{tabular}{||l|l||c|c||}
\hline
Observable & Exp. bound on  $|\delta|$ & $|\delta|$ in $SU(5)_{RN}$ \\
\hline
 ${\rm BR}(\mu\to e\gamma)$ & $|\delta^{L}_{\mu e}|< 3\times 10^{-4}$  & $\sim 3\times10^{-4}$ \\
\hline
 ${\rm BR}(\mu\to e\gamma)$ & $|\delta^{L}_{\mu\tau}\delta^{L}_{\tau e}|< 10^{-3}$ & $\sim 10^{-6}$ \\
\hline
 ${\rm BR}(\mu\to e\gamma)$ & $|\delta^{L}_{\mu\tau}\delta^{R}_{\tau e}|< 10^{-3}$
 & $\sim 10^{-5}$ \\
\hline
 ${\rm BR}(\tau\to e\gamma)$ & $|\delta^{L}_{\tau e}|< 6\times 10^{-2}$  & $\sim 3\times10^{-4}$\\
\hline
 ${\rm BR}(\tau\to\mu\gamma)$ & $|\delta^{L}_{\tau \mu }|< 4\times 10^{-2}$  & $\sim 4\times10^{-3}$\\
\hline
\end{tabular}
\end{center}
\caption{\label{lfvtable2}
Bounds on the effective LFV couplings $\delta_{l_i l_j}^{L}$ from the current experimental bounds
on the radiative LFV decays of $\tau$ and $\mu$ leptons (see Table~\ref{lfvtable}) by setting
$\Delta a^{\rm SUSY}_{\mu}=3 \times 10^{-9}$. The expectations for the $\delta_{l_i l_j}^{L}$'s within the
$SU(5)_{RN}$ scenario with hierarchical right-handed neutrinos are reported in the last column and they correspond
to the reference values ${M}^{\rm (r)}_{3}= 10^{13}\, {\rm GeV}$ and ${U}^{\rm (r)}_{e3}=0.1$. The predictions
for $\delta_{\mu e}^{L}$ and $\delta_{\tau e}^{L}$ scale with $U_{e3}$ as $(U_{e3}/{U}^{\rm (r)}_{e3})$
while all the $\delta_{l_i l_j}^{L}$'s scale with $M_{3}$ as $(M_{3}/{M}^{\rm (r)}_{3})$.
Improving the experimental resolutions on ${\rm BR}(\mu\to e\gamma)$, the bounds on $\delta^{L}_{\mu e}$
and $\delta^{L}_{\mu\tau}\delta^{R(L)}_{\tau e}$ reported in this Table will scale as
$[{\rm BR}(\mu\to e\gamma)_{exp}/1.2\times 10^{-11}]^{1/2}$. The scaling properties for the other
flavor transitions are obtained in the same way.}

\end{table}
\section{Leptonic EDMs}
\label{sec:EDMs}

Within a SUSY framework, CP-violating sources are naturally induced by the soft SUSY breaking
terms through {\it i)} flavor conserving $F$-terms (such as the $B\mu$ parameter in the Higgs 
potential or the $A$ terms for trilinear scalar couplings)~\cite{pospelov} and {\it ii)}
flavor-violating $D$-terms (such as the squark and slepton mass terms)~\cite{HNP_EDM}. It seems
quite likely that the two categories {\it i)} and {\it ii)} of CP violation are controlled by
different physical mechanisms, thus, they may be distinguished and discussed independently.

In the case {\it i)}, it is always possible to choose a basis where only the $\mu$ and $A$
parameters remain complex~\cite{pospelov}. The CP-violating phases generally lead to
large electron and neutron EDMs as they arise already at the one-loop level. For example,
when $t_{\beta}=10$, $\tilde{m}=300\,{\rm GeV}$, $d_e\sim 6\times 10^{-25}(\sin\theta_\mu + 10^{-2}\sin\theta_A)\,e\,$cm.

In the case {\it ii)}, the leptonic EDMs induced by {\it flavor dependent} phases (flavored EDMs) 
read~\cite{HNP_EDM}
\beq
\frac{d_{\ell}}{e}\simeq
-\frac{\alpha_Y}{4\pi}\bigg(\frac{m_\tau}{\tilde{m}^{2}}\bigg)\,\tgb~
\frac{{\rm Im}\left(\delta^{R}_{\ell\tau}\delta^{L}_{\tau\ell}\right)}{30}\,,
\label{edm_flavor}
\eeq
where a common SUSY mass $\tilde{m}$ has been assumed. If $t_{\beta}=10$ and 
$\tilde{m}=300\,{\rm GeV}$, it turns out that
$d_e\sim 10^{-22}\times{\rm Im}(\delta^{R}_{e\tau}\delta^{L}_{\tau e})\,e\,$cm.

One of the most peculiar features disentangling the EDMs as induced by {\it flavor blind}
or {\it flavor dependent} phases regards their scaling properties with different leptons.
In particular,
\bea
\frac{d_e}{d_{\mu}}\!\!&=&\!\!\frac{m_e}{m_{\mu}}\quad\quad\quad\quad\quad\,\,\, {\it flavor\, blind}\, 
{\rm phases},
\nonumber\\
\frac{d_e}{d_{\mu}}\!\!&=&\!\!\frac{{\rm Im}(\delta^{R}_{e\tau}\delta^{L}_{\tau e})}
{{\rm Im}(\delta^{R}_{\mu\tau}\delta^{L}_{\tau\mu})}\quad\quad {\it flavor\, dependent}\, {\rm phases}\,.
\eea
In the case of {\it flavor blind} phases, the current bound $d_{e}<1.7\times 10^{-27}\,e\,$cm \cite{Amsler:2008zzb}
implies that $d_{\mu}\lesssim 3.5 \times 10^{-25}\,e\,$cm. On the contrary, in presence of
{\it flavor dependent} phases, the leptonic EDMs typically violate the naive scaling and
values for $d_{\mu}> 2 \times 10^{-25}\,e\,$cm are still allowed.

Moreover, when the EDMs are generated by {\it flavor blind} phases, they are completely unrelated
to LFV transitions (although correlations with CP and flavor violating transitions in the $B$-meson
systems are still possible~\cite{ABP}). By contrast, the flavored EDMs are closely related to LFV
processes as $\ell_i\to \ell_j\gamma$ since they are both generated by LFV effects and they arise
from similar dipole operators. If the EDMs and LFV processes will be observed, their correlation
will provide a precious tool to disentangle the LFV source responsible for LFV transitions.

Actually, if ${\rm BR}(\mu\to e\gamma)$ is dominated by the term $\delta^R_{\mu\tau}\delta^L_{\tau e}$
and/or $\delta^L_{\mu\tau}\delta^R_{\tau e}$, we get
\beq
 d_e
  \simeq
  \frac{d_\mu}{x_R x_L}
  \simeq
  2\times 10^{-26}\,e\,{\rm cm}\, 
  \sqrt{\frac{{\rm BR}(\mu \to e\gamma)/10^{-11}}{x_R^2+x_L^2}}\,,
\label{dedmu}
\eeq
where $x_{R/L}=|\delta_{\mu\tau}^{R/L}/\delta_{e\tau}^{R/L}|$ and maximum phases have been assumed.
Eq.~(\ref{dedmu}) implies the following constraint for the flavored leptonic EDMs,
\beq
d_e d_\mu \lsim \left(1.6 \times 10^{-26}\,e\,{\rm cm}\right)^2 \times
\frac{{\rm BR}(\mu \to e\gamma)}{10^{-11}}~.
\label{dedmubound}
\eeq
The bound of Eq.~(\ref{dedmubound}) arises when ${\rm BR}(\mu\to e\gamma)$ is generated by
the combination of left- and right-handed slepton mixing, it doesn't depend on the details
of the SUSY spectrum and it is saturated when $x_R=x_L$.

We observe that, within the $SU(5)_{RN}$ scenario, $d_e$ grows linearly with $U_{e3}$ since
$d_e\sim\delta^{R}_{e\tau}\delta^{L}_{\tau e}$ with $\delta^{L}_{\tau e}\sim U_{e3}$ while
$d_{\mu}$ does not depend on $U_{e3}$. As a result, it turns out that 
\beq
\left(\frac{d_e}{d_{\mu}}\right)_{{SU(5)}_{RN}}
\simeq \frac{U_{e3}V_{td}}{U_{\mu 3}V_{ts}}\,.
\eeq
In the pure SUSY see-saw model, the leptonic EDMs are highly suppressed if the right-handed
neutrino masses are degenerate, similarly to the quark sector. In contrast, if the right-handed
neutrinos are not degenerate, the predictions for the EDMs might be significantly enhanced by
means of threshold corrections to the SUSY breaking terms~\cite{edminseesaw}. The EDMs are
sensitive to the Dirac and Majorana phases in the MNS matrix as well as to the phases in $R$.
However, they still remain well below any future (realistic) experimental resolution. Indeed,
we have explicitly checked that, after imposing the current experimental bound on
${\rm BR}(\mu\to e\gamma)$, we end up with contributions to the electron EDM of order
$d_e \!\lsim \! 10^{-34}~e$cm, irrespective to the details of the heavy/light neutrino sectors.
On the contrary, when the see-saw mechanism is embedded in a SUSY GUT scheme, as $SU(5)_{RN}$,
$d_e$ may naturally saturate its current experimental upper bound.
Hence, any experimental evidence for the leptonic EDMs at the upcoming experiments could
naturally points towards a SUSY GUT framework with an underlying see-saw mechanism specially
if $\mu\to e\gamma$ would also be observed at the MEG experiment.

Interestingly enough, the synergy of apparently unrelated low-energy experiments, as the leptonic
EDMs and LFV processes (like $\mu\to e\gamma$), represents a powerful tool to shed light on the
underlying NP theory that is at work.

\section{The P-odd asymmetry in $\mu^{+}\to e^{+}\gamma$}
\label{sec:ameg}

In case LFV processes as $\ell_i\to \ell_j\gamma$ will be observed at the upcoming experiments,
a crucial question would be to understand which is the kind of LFV source responsible for such
a NP signal. In this respect, a very useful tool will be provided by the asymmetries defined
by means of initial muon polarization.
Experimentally, polarized positive muons are available by the surface muon method because muons
emitted from $\pi^{+}$'s stopped at target surface are $100\%$ polarized in the direction opposite
to the muon momentum. Interestingly enough, in Ref.~\cite{96KunoOkada} it has been shown that the
muon polarization is useful to suppress the background processes in the $\mu^{+}\to e^{+}\gamma$
search. As for the signal distribution of $\mu^{+}\to e^{+}\gamma$, the angular distribution
with respect to the muon polarization can distinguish between $\mu^{+}\rightarrow e_{L}^{+}\gamma$
and $\mu^{+}\to e_{R}^{+}\gamma$. In particular, one can define the P-odd asymmetry
$A(\mu^{+}\to e^{+}\gamma)$ as~\cite{Okada:1999zk}
\beq
\ A(\mu^{+}\to e^{+}\gamma) = \frac{|A_{L}|^{2}-|A_{R}|^{2}}{|A_{L}|^{2}+|A_{R}|^{2}}\,.
\label{Eq:asymmetry}
\eeq
As we will show, the knowledge of $A(\mu^{+}\to e^{+}\gamma)$ will represent a powerful tool to
shed light on the nature of the LFV sources, in particular to disentangle whether an underlying
SUSY GUT theory is at work or not.

In fact, a pure (non-GUT) SUSY see-saw predicts $A(\mu^+\to e^+\gamma)=+1$ to a very good accuracy,
as the largely dominant amplitude is $A^{\mu e}_L \sim \delta^L_{\mu e}$ (in fact it turns out
that $A^{\mu e}_R \sim (m_e/m_{\mu})\times A^{\mu e}_L$). Thus, any experimental evidence
departing from this expectation would likely support the idea of a SUSY see-saw model embedded
in GUT scenarios where, in addition to $A^{\mu e}_L$, a sizable amplitude
$A^{\mu e}_R\sim\delta^L_{\mu \tau}\delta^R_{\tau e}$ is also generated. Should this happen, we
would also expect large values for ${\rm BR}(\tau\to\mu\gamma)$ arising from $\delta^{L}_{\mu\tau}$.

\section{Numerical analysis}
\label{sec:num_analysis}

In this section, we present the numerical results relative to the observables discussed in the
previous sections both in the low-energy (model independent) approach and in the $SU(5)_{RN}$
model described in previous section.

\begin{figure}
\includegraphics[scale=0.65]{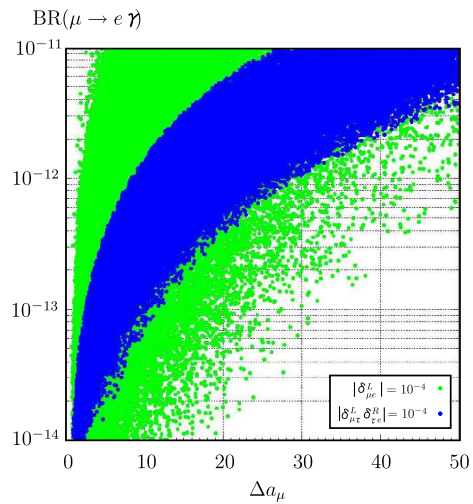}
\caption{${\rm BR}(\mu\to e\gamma)$ vs the SUSY contribution to the muon anomalous magnetic moment
$\Delta a_\mu^{\rm SUSY}$. The plot has been obtained by means of a scan over the following SUSY 
parameter space: $3<\tan\beta<50$, $(m_{\tilde{\ell}},\mu, M_{\tilde{W}}=2M_{\tilde{B}})\leq 1$~TeV.
Blue points correspond to the case where ${\rm BR}(\mu\to e\gamma)$ is generated by the only
$\delta^L_{\mu e}$ MI (we set $|\delta^L_{\mu e}|=10^{-4}$), while green points refer to the case
where ${\rm BR}(\mu\to e\gamma)$ is generated by the only $\delta^L_{\mu \tau}\delta^R_{\tau e}$ MI
(we set $(|\delta^L_{\mu \tau}\delta^R_{\tau e}|=10^{-4})$). For different values of $|\delta^L_{\mu e}|$
and $|\delta^L_{\mu \tau}\delta^R_{\tau e}|$, ${\rm BR}(\mu\to e\gamma)$ scales as
$(|\delta^L_{\mu e}|/10^{-4})^2$ and $(|\delta^L_{\mu \tau}\delta^R_{\tau e}|/10^{-4})^2$, respectively.}
\label{Fig:gm2LL}
\end{figure}

Starting with the model independent analysis, in Fig.~\ref{Fig:gm2LL}, we show the predictions
for ${\rm BR}(\mu\to e\gamma)$ and $\Delta a_\mu^{\rm SUSY}$ as obtained by means of a scan over
the SUSY parameters $3<\tan\beta<50$, $(m_{\tilde{\ell}},\mu, M_{\tilde{W}}=2M_{\tilde{B}})\leq 1$~TeV,
assuming a common slepton mass $m_{\tilde{\ell}}$.

Blue points refer to the case where ${\rm BR}(\mu\to e\gamma)$ is generated only by $\delta^L_{\mu e}$;
the quite strong correlation between ${\rm BR}(\mu\to e\gamma)$ and $\Delta a_\mu^{\rm SUSY}$ does not
change significantly if $\delta^L_{\mu\tau}\delta^L_{\tau e}$ also contributes to 
${\rm BR}(\mu\to e\gamma)$. Green points refer to the case where ${\rm BR}(\mu\to e\gamma)$ is generated
only by $\delta^L_{\mu \tau}\delta^R_{\tau e}$; now, the correlation between ${\rm BR}(\mu\to e\gamma)$
and $\Delta a_\mu^{\rm SUSY}$ is rather loose with respect to the previous case. This behavior
can be understood remembering that ${\rm BR}(\mu\to e\gamma)$ is induced now only by the $U(1)_Y$
interactions by means of the pure Bino exchange. Still, some useful information may be extracted
from Fig.~\ref{Fig:gm2LL}: the explanation of the muon $(g-2)$ anomaly through SUSY effects implies
a lower bound for ${\rm BR}(\mu\to e\gamma)$ which clearly depends on the size of the LFV source.

%

\begin{figure}
\includegraphics[scale=0.65]{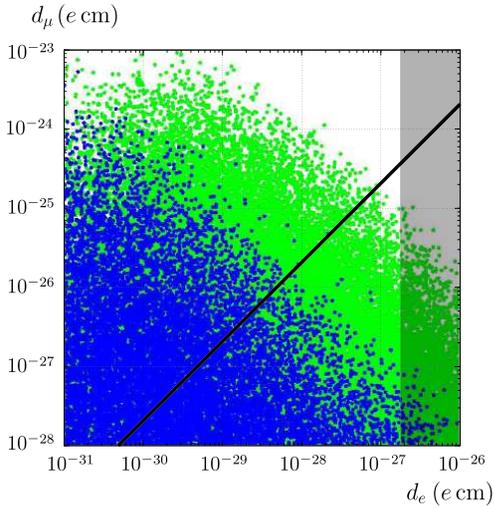}
\caption{Model independent correlation between $d_e$ vs $d_{\mu}$ with the same input
parameters of Fig.~\ref{Fig:gm2LL} and varying
$10^{-5}<(\delta^{L,R}_{e\mu},\delta^{L,R}_{e\tau},\delta^{L,R}_{\mu\tau})<1$.
The current experimental constraints on ${\rm BR}(\tau\to\mu\gamma)$ and
${\rm BR}(\tau\to e\gamma)$ have been imposed. Moreover, the green and blue points
correspond to ${\rm BR}(\mu\to e\gamma)<(10^{-11},10^{-13})$, respectively.
The black line corresponds to the naive scaling $d_e/d_{\mu}=m_e/m_{\mu}$.
The grey region is excluded by the current experimental upper bound on $d_e$.
}
\label{Fig:dedmu}
\end{figure}

In Fig.~\ref{Fig:dedmu}, we show the allowed regions for $d_e$ and $d_{\mu}$ compatible with the
current upper bounds on ${\rm BR}(\tau\to e\gamma)$ and ${\rm BR}(\tau\to\mu\gamma)$; the green
(blue) region corresponds to ${\rm BR}(\mu\to e\gamma)\leq 10^{-11} (10^{-13})$. The plot has been
obtained through a scan over the same input parameters of Fig.~\ref{Fig:gm2LL} with the addition of $10^{-5}<(\delta^{L,R}_{e\mu},\delta^{L,R}_{e\tau},\delta^{L,R}_{\mu\tau})<1$, and the LFV sources are
treated in a model-independent way allowing, in particular, for maximum CP-violating phases.
The black line in Fig.~\ref{Fig:dedmu} corresponds to the naive scaling of the leptonic EDMs, {\it i.e.} $d_e/d_{\mu}=m_e/m_{\mu}$, as it would happen if the EDMs were generated by flavor blind phases.

%


We now pass to the numerical analysis relative to the $SU(5)_{RN}$ model.
In general, since SUSY GUT models present a rich flavor structure, many flavor-violating phenomena~\cite{gut_FCNC} as well as leptonic and hadronic (C)EDMs are generated~\cite{gutedm}.
Moreover, since within SUSY GUTs leptons and quarks sit into same multiplets, the flavor violation
in the squark and slepton sectors may be correlated~\cite{gut_FCNC}. However, in this paper, we
focus only on the $SU(5)_{RN}$ predictions for the leptonic sector, although the hadronic processes
are systematically taken into account to constrain the SUSY parameter space.

In the following, we assume the gravity mediated mechanism for the SUSY breaking terms and we
take $M_X=2.4\times 10^{18}$~GeV.

\begin{figure}
\includegraphics[scale=0.65]{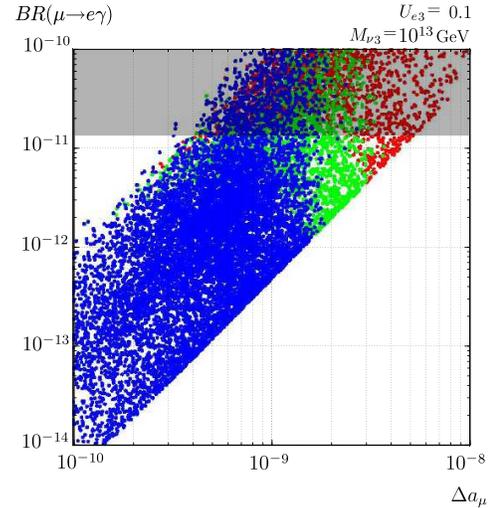}
\caption{
${\rm BR}(\mu\to e\gamma)$ vs $\Delta a^{\rm SUSY}_{\mu}$ in the $SU(5)_{RN}$ model assuming a hierarchical
spectrum for both light and heavy neutrinos, $m_{\nu_3}=0.05{\rm eV}$, $M_{3}=10^{13}\,{\rm GeV}$
and $U_{e3}=0.1$. The plot has been obtained varying the SUSY parameters in the following ranges:
$100\,{\rm GeV}<m_0,M_{1/2}<1\,{\rm TeV}$, $|A_0|<3m_0$, $3<\tan\beta<50$ and $\mu>0$. Green (blue)
points satisfy the constraints from ${\rm BR}(B\to X_s\gamma)$ at the $99\%$ C.L. ($90\%$ C.L.) limit.
The grey region is excluded by the current experimental upper bound on ${\rm BR}(\mu\to e\gamma)$.}
\label{megm2_su5}
\end{figure}

In Fig.~\ref{megm2_su5}, we show the predictions for ${\rm BR}(\mu\to e\gamma)$ vs $\Delta a^{\rm SUSY}_{\mu}$
assuming $m_{\nu_3}=0.05{\rm eV}$, $M_{3}=10^{13}\,{\rm GeV}$ and $U_{e3}=0.1$ and varying the SUSY parameters in the ranges $m_0,M_{1/2}<1\,{\rm TeV}$, $|A_0|<3m_0$, $3<\tan\beta<50$ and $\mu>0$.

The blue (green) points satisfy the constraints from ${\rm BR}(B\to X_s\gamma)$~\cite{Amsler:2008zzb} 
at the $99\%$ $(90\%)$ C.L. limit~\footnote{We have evaluated ${\rm BR}(B\to X_s\gamma)$ including the 
SM effects at the NNLO~\cite{misiak} and the NP contributions at the LO in this paper.}, while the red 
ones do not.
As shown in Fig.~\ref{megm2_su5}, sizable SUSY effects to the muon $(g-2)$, at the level of
$\Delta a^{\rm SUSY}_{\mu}\sim 10^{-9}$, lead to values for ${\rm BR}(\mu\to e\gamma)$ well within
the MEG reach for natural values of the neutrino parameters $M_{3}$ and $U_{e3}$. Moreover, we
note that the constraint from ${\rm BR}(B\to X_s\gamma)$ at the $90\%$ ($99\%$) C.L. allows SUSY
contributions to the muon $(g-2)$ as large as $\Delta a^{\rm SUSY}_{\mu}\lesssim 1(2)\times 10^{-9}$.
%

\begin{figure}
\includegraphics[scale=0.65]{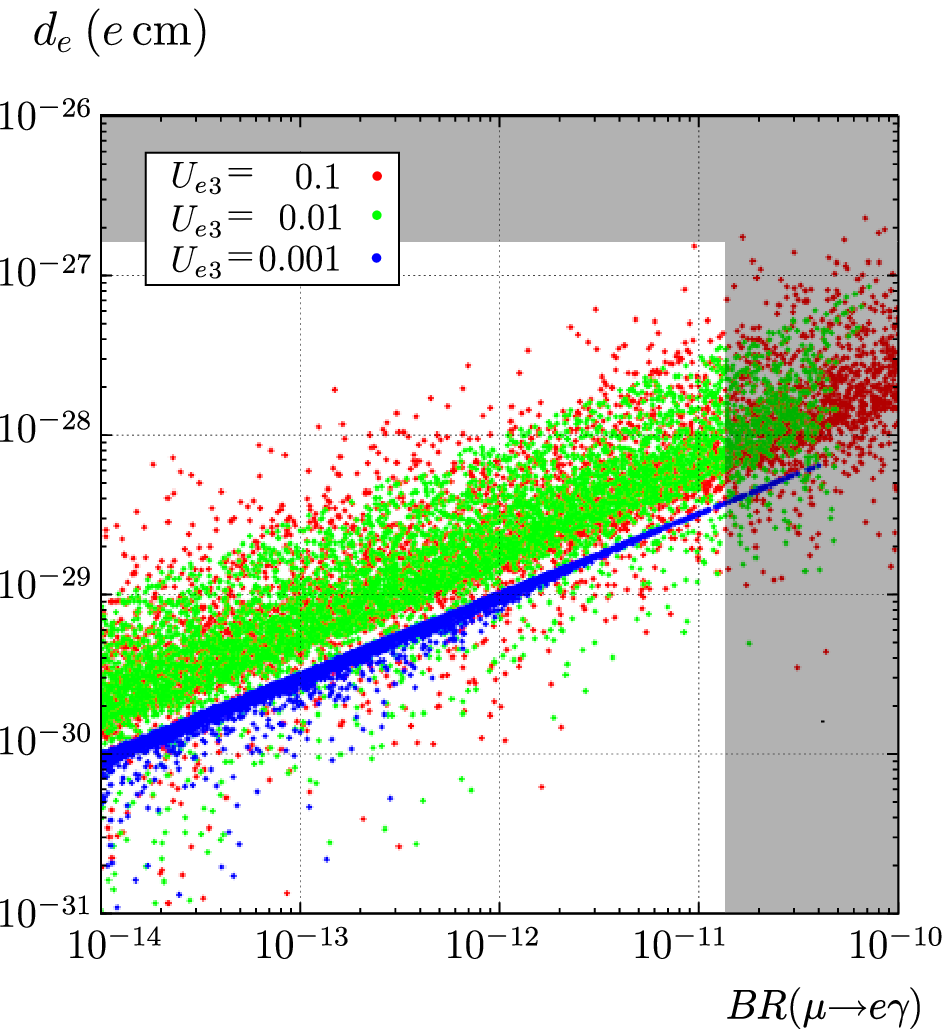}
\includegraphics[scale=0.65]{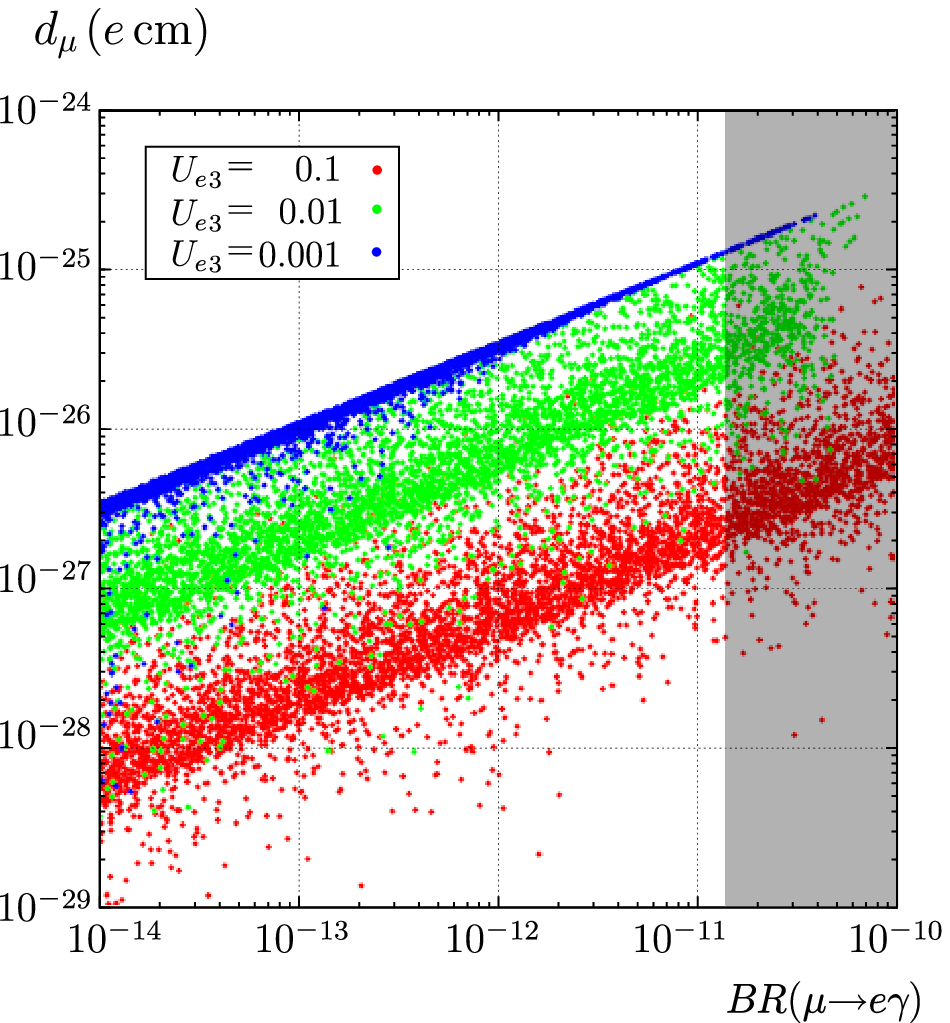}
\caption{In the upper (lower) plot we show the electron (muon) EDM $d_e$ ($d_{\mu}$) vs
${\rm BR}(\mu\to e\gamma)$ in the $SU(5)_{RN}$ model assuming maximum CP-violating phases.
The input parameters are given as $m_0,M_{1/2}<1\,{\rm TeV}$, $|A_0|<3m_0$, $3<\tan\beta<50$
and $\mu>0$. For the neutrino sectors, we assume a hierarchical spectrum for both light and 
heavy neutrinos and we take $m_{\nu_3}=0.05{\rm eV}$, $10^{10}<M_{3}<10^{15}\,{\rm GeV}$.
The grey regions are excluded by the current experimental upper bounds on ${\rm BR}(\mu\to e\gamma)$ 
and $d_e$.}
\label{Fig:meg_dl}
\end{figure}

In Fig.~\ref{Fig:meg_dl}, we show the electron and muon EDMs vs ${\rm BR}(\mu\to e\gamma)$
assuming maximum CP-violating phases. We vary the input parameters as $m_0,M_{1/2}<1\,{\rm TeV}$,
$|A_0|<3m_0$, $3<\tan\beta<50$ and $\mu>0$; we also take $m_{\nu_3}=0.05{\rm eV}$,
$10^{10}<M_{3}<10^{15}{\rm GeV}$ and we consider three different values for $U_{e3}=0.1,0.01,0.001$.
The attained values by $d_e$ and $d_{\mu}$, compatible with the current experimental bound on
${\rm BR}(\ell_i\to \ell_j\gamma)$, are well within the expected future experimental sensitivities
for $d_e$, at least. 
It is noteworthy that, even in the pessimistic case in which $\mu\to e\gamma$
will not be observed at the MEG experiment (at the level of ${\rm BR}(\mu\to e\gamma)\lesssim 10^{-13}$), the predictions for $d_e$ are still typically well above the level of $10^{-31} e~{\rm cm}$.

Besides the running MEG experiment, also other experiments, i.e. Mu2e at Fermilab~\cite{mu2e} and
COMET at J-parc~\cite{comet}, looking for $\mu$-$e$ conversion in nuclei with expected sensitivities of 
order $10^{-(16-17)}$, are planed. These sensitivities would indirectly probe ${\rm BR}(\mu\to e\gamma)$
at the level of ${\rm BR}(\mu\to e\gamma)\lesssim 10^{-14}$ (see Eq.(\ref{eq:dipole})).
Furthermore, the PRISM/PRIME experiment, in which a very intensive pulsed beam is produced by the FFAG
muon storage ring, is also planed and its ultimate sensitivity to $\mu$-$e$ conversion in nuclei should
reach the $10^{-(18-19)}$ level~\cite{comet}. Thus, $\mu$--$e$ conversion experiments and the electron
EDM would represent the most promising and powerful tool to probe the $SU(5)_{RN}$ model after the MEG experiment.

We remind that when $U_{e3}$ is very small, $d_e$, $d_{\mu}$ and ${\rm BR}(\mu\to e\gamma)$ turn out
to be highly correlated, as they are generated by very similar Bino induced diagrams; looking at Fig.~\ref{Fig:meg_dl}, this correlation is evident in the case of blue points, corresponding to
$U_{e3}=10^{-3}$. In the scenario with a negligibly small $U_{e3}\leq 10^{-3}$, both ${\rm BR}(\tau\to\mu\gamma)$ and $d_{\mu}$ assume their maximum values as the constraints from ${\rm BR}(\mu\to e\gamma)$ are quite
relaxed in this case.

\begin{figure}
\includegraphics[scale=0.65]{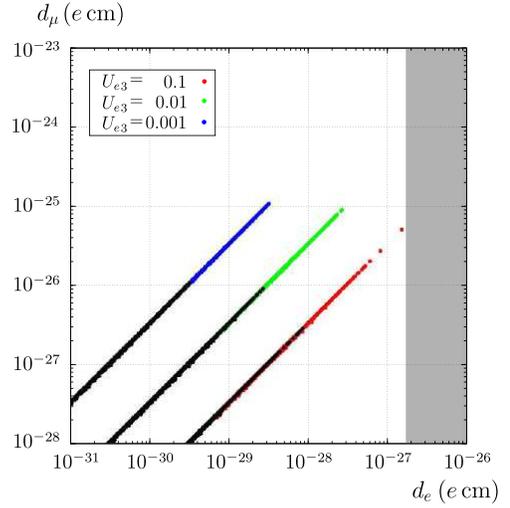}
\caption{Electron EDM $d_e$ vs muon EDM $d_{\mu}$ in the $SU(5)_{RN}$ model assuming
maximum CP-violating phases. The input parameters are given as in Fig.~\ref{Fig:meg_dl}
and the current constraints from ${\rm BR}(\ell_i\to \ell_j\gamma)$ have been imposed.
The black dots correspond to ${\rm BR}(\mu\to e\gamma)\leq 10^{-13}$.
The grey region is excluded by the current experimental upper bound on $d_e$.}
\label{Fig:dmu_de}
\end{figure}

In Fig.~\ref{Fig:dmu_de}, we show the correlation between $d_e$ vs $d_{\mu}$ assuming maximum
CP-violating phases and the same input parameters as in Fig.~\ref{Fig:meg_dl}.
As shown by the Eq.~(\ref{dedmubound}), the flavored leptonic EDMs are bounded by the experimental
limit on ${\rm BR}(\mu\to e\gamma)$. The dots excluded at the levels of
${\rm BR}(\mu\to e\gamma)<10^{-11}$ and ${\rm BR}(\mu\to e\gamma)<10^{-13}$ are also indicated
in Fig.~\ref{Fig:dmu_de}.
%
%

\begin{figure}
\includegraphics[scale=0.65]{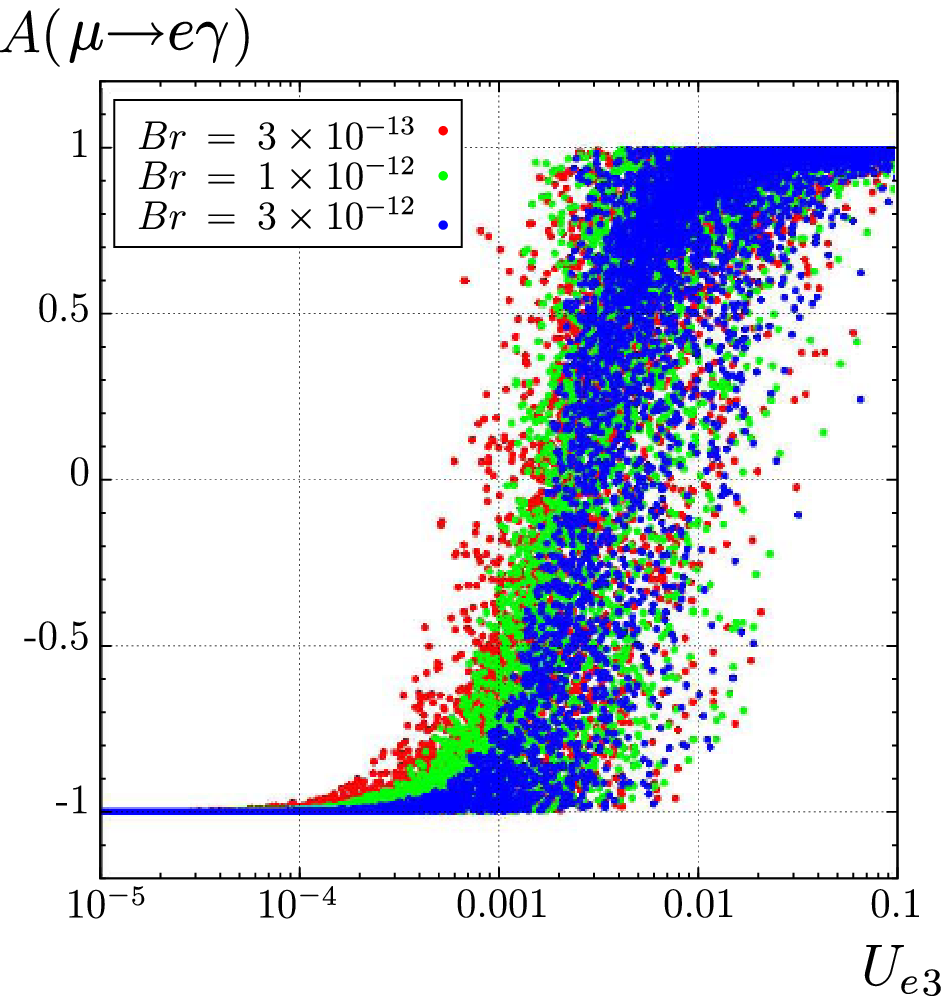}
\includegraphics[scale=0.65]{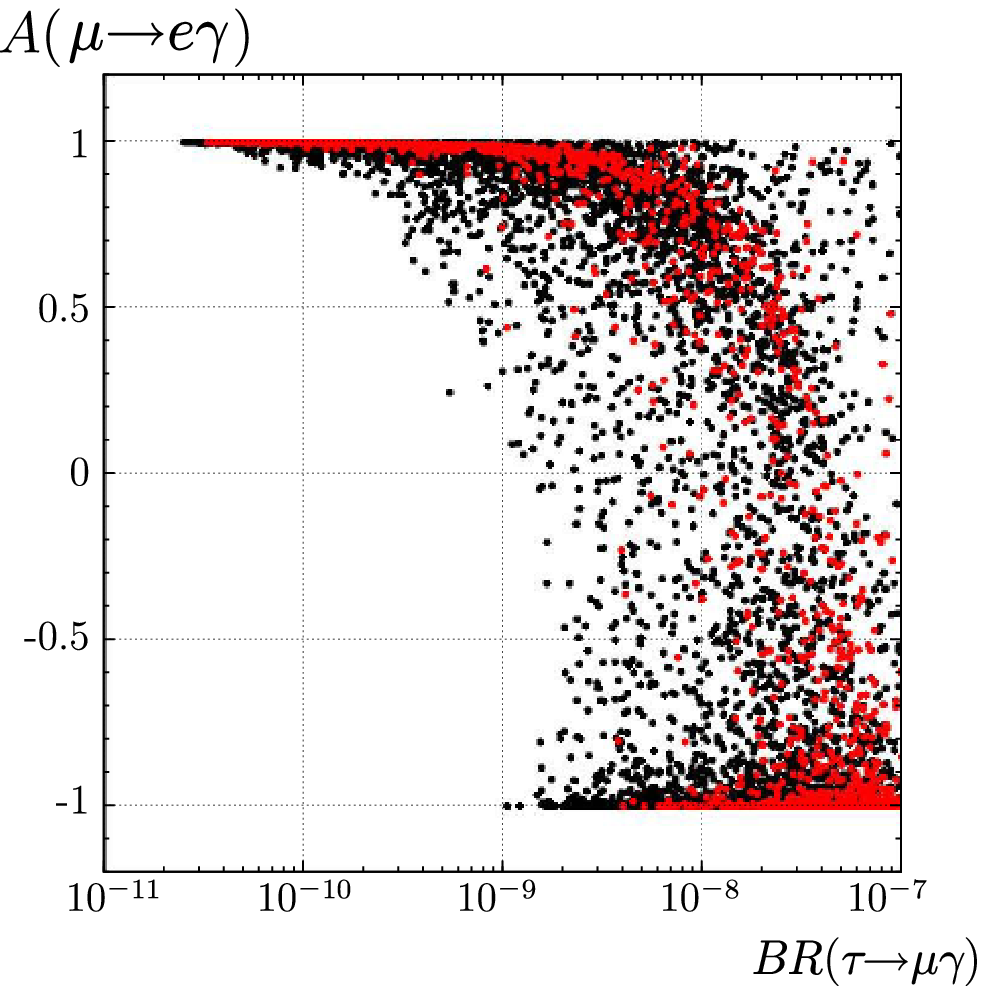}
\caption{Upper plot: P-odd asymmetry in $\mu^{+}\to e^{+}\gamma$, $A(\mu^+\to e^+\gamma)$, vs
$U_{e3}$ in the $SU(5)_{RN}$ model for three different values of ${\rm BR}(\mu\to e\gamma)
=(3,1,0.3)\times 10^{-12}$. Lower plot: $A(\mu^+\to e^+\gamma)$ vs ${\rm BR}(\tau\to \mu\gamma)$
assuming ${\rm BR}(\mu\to e\gamma)=3\times 10^{-12}$. Both plots have been obtained by means
of a scan of the input parameters $m_0,M_{1/2}<1\,{\rm TeV}$, $|A_0|<3m_0$, $3<\tan\beta<50$
and $\mu>0$. For the neutrino sectors, we have assumed a hierarchical spectrum for both light
and heavy neutrinos and we take $m_{\nu_3}=0.05{\rm eV}$, $10^{10}<M_{3}<10^{15}\,{\rm GeV}$
and $10^{-5}\leq U_{e3}\leq 0.1$. All the points of both plots satisfy the constraints from
$b\to s\gamma$ at the 99\% C.L. limit and $m_{h^0}>111.4$~GeV. Red points in the lower plot also
satisfy $\Delta a^{\rm SUSY}_{\mu}\geq 1\times 10^{-9}$.}
\label{Fig:Ue3_Ameg}
\end{figure}

In the upper plot of Fig.~\ref{Fig:Ue3_Ameg}, we show the values attained by the P-odd asymmetry
$A(\mu^+\to e^+\gamma)$ given in Eq.~(\ref{Eq:asymmetry}) as a function of $U_{e3}$ for three
different values of ${\rm BR}(\mu\to e\gamma)=(3, 1, 0.3)\times 10^{-12}$, corresponding to the
green, red and blue bands of Fig.~\ref{Fig:Ue3_Ameg}, respectively.

The plot has been obtained in the following way: we have performed a scan over the input
parameters $m_0,M_{1/2}<1\,{\rm TeV}$, $|A_0|<3m_0$, $3<\tan\beta<50$ (we set $\mu>0$) and $10^{10}<M_{3}<10^{15}\,{\rm GeV}$ (we set $m_{\nu_3}=0.05{\rm eV}$). Then, after imposing
all the existing constraints arising from flavor observables (both in the leptonic and hadronic
sectors), direct searches as well as from theoretical constraints, we have selected all the
sets of input parameters producing a same value for ${\rm BR}(\mu\to e\gamma)$; in particular
we have considered the three cases ${\rm BR}(\mu\to e\gamma)=(3, 1, 0.3)\times 10^{-12}$.

We note that when the parameter $U_{e3}$ is large ($U_{e3}\sim 0.1$), the ${\rm BR}(\mu\to e\gamma)$
is almost determined by the amplitude $A_L\sim\delta^{\mu e}_{L}\sim U_{e3}$ and we expect
$A(\mu^+\to e^+\gamma)\sim +1$, as it is confirmed numerically by the Fig.~\ref{Fig:Ue3_Ameg}.
In contrast, when $U_{e3}$ is very small ($U_{e3}\lesssim 10^{-4}$), the ${\rm BR}(\mu\to e\gamma)$
is dominated by the amplitude $A_R\sim\delta^{\mu\tau}_{L}\delta^{\tau e}_{R}$ and $A(\mu^+\to e^+\gamma)$
approaches to $-1$, as shown by the Fig.~\ref{Fig:Ue3_Ameg}.
When $U_{e3}$ is neither very close to $0.1$ nor to zero, we expect $A(\mu^+\to e^+\gamma)$ in the
range $A(\mu^+\to e^+\gamma)\in(-1,+1)$. It is noteworthy to observe
that already for $U_{e3}$ values not so far from $0.1$, $A(\mu^+\to e^+\gamma)$ can depart sizably
from $A(\mu^+\to e^+\gamma)=+1$.


In the lower plot of Fig.~\ref{Fig:Ue3_Ameg}, we show the correlation between $A(\mu^+\to e^+\gamma)$ and
${\rm BR}(\tau\to \mu\gamma)$ assuming an experimental evidence for $\mu\to e\gamma$ at the level of
${\rm BR}(\mu\to e\gamma)=3\times 10^{-12}$. It is found that a sizable departure
from the value $A(\mu^+\to e^+\gamma)=+1$ would most likely imply a lower bound for $\tau\to \mu\gamma$.
In Fig.~\ref{Fig:Ue3_Ameg} it turns out that ${\rm BR}(\tau\to \mu\gamma)\gtrsim 10^{-9}$ and this is
specially true if we also require an explanation of the muon $(g-2)$ anomaly in terms of SUSY effect,
as shown by the red points in the lower plot of Fig.~\ref{Fig:Ue3_Ameg}, corresponding to
$\Delta a^{\rm SUSY}_{\mu}\geq 1\times 10^{-9}$. If we assume values for ${\rm BR}(\mu\to e\gamma)$ smaller than
${\rm BR}(\mu\to e\gamma)=3\times 10^{-12}$, the corresponding predictions for ${\rm BR}(\tau\to \mu\gamma)$
will decrease of the same factor as ${\rm BR}(\mu\to e\gamma)$.

%

\begin{figure}
\includegraphics[scale=0.65]{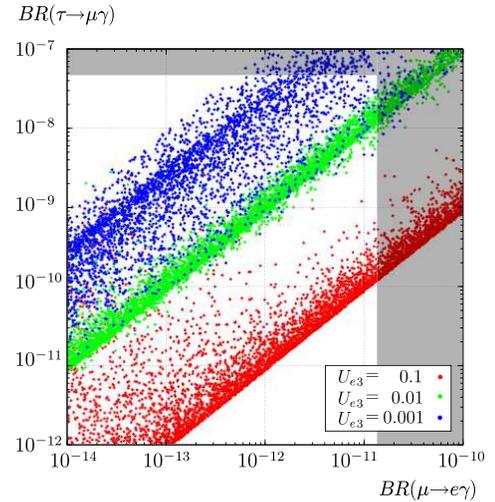}
\caption{${\rm BR}(\mu\to e\gamma)$ vs ${\rm BR}(\tau\to\mu\gamma)$ in the $SU(5)_{RN}$ model.
The plot has been obtained by means of a scan over the same input parameters of Fig.~\ref{Fig:meg_dl}.
The grey regions are excluded by the current experimental upper bounds on ${\rm BR}(\mu\to e\gamma)$
and ${\rm BR}(\tau\to \mu\gamma)$.}
\label{Fig:meg_tmg}
\end{figure}

In Fig.~\ref{Fig:meg_tmg}, we show the correlation between ${\rm BR}(\mu\to e\gamma)$ and
${\rm BR}(\tau\to\mu\gamma)$ for three different values of $U_{e3}=(0.001,0.01,0.1)$. We recall
that while ${\rm BR}(\tau\to\mu\gamma)$ is not sensitive to $U_{e3}$, ${\rm BR}(\mu\to e\gamma)$
crucially depends on $U_{e3}$. As shown in Fig.~\ref{Fig:meg_tmg}, if $U_{e3}=0.1$, namely if
$U_{e3}$ is close to its current experimental upper bound, the current bound ${\rm BR}(\mu\to e\gamma)
\lesssim 10^{-11}$ already implies that ${\rm BR}(\tau\to\mu\gamma)\lesssim 10^{-9}$, a level that
is most probably beyond the reach of the Super $B$ factories. In such a case, it is clear that 
$\mu\to e\gamma$ would represent the golden channel where to look for SUSY LFV signals given the
expected experimental resolutions at the running MEG experiment ${\rm BR}(\mu\to e\gamma)\lesssim 10^{-13}$. 
In contrast, as shown in Fig.~\ref{Fig:meg_tmg}, if $U_{e3}$ will turn out to be smaller than 
$U_{e3}=0.1$, there are regions where both MEG and the Super $B$ factories are expected to detect 
LFV signals. In the extreme case where $U_{e3}$ is very small, say $U_{e3}\lesssim 10^{-3}$, 
$\tau\to\mu\gamma$ could still lie well within the Super $B$ factories reach while
${\rm BR}(\mu\to e\gamma)$ could result too small to be seen at the MEG experiment.

Hence, we want to stress here that $\mu\to e\gamma$ and $\tau\to\mu\gamma$ are very important and
complementary probes of LFV effects arising in SUSY theories.

\begin{figure}[t]
\includegraphics[scale=1.1]{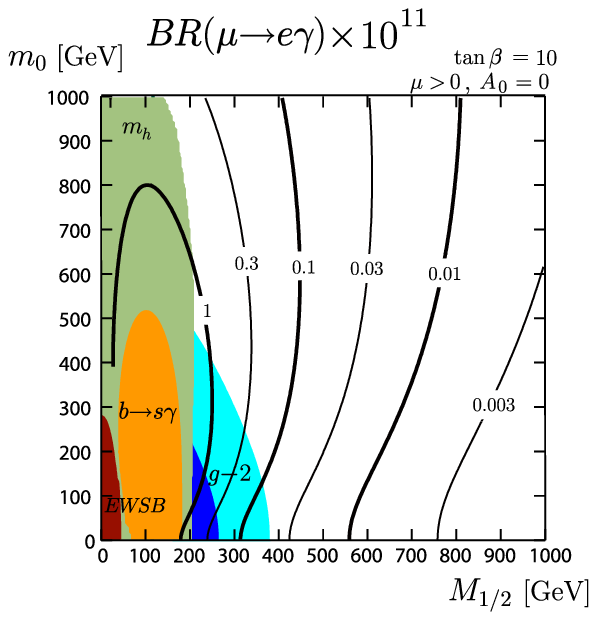}
\includegraphics[scale=1.1]{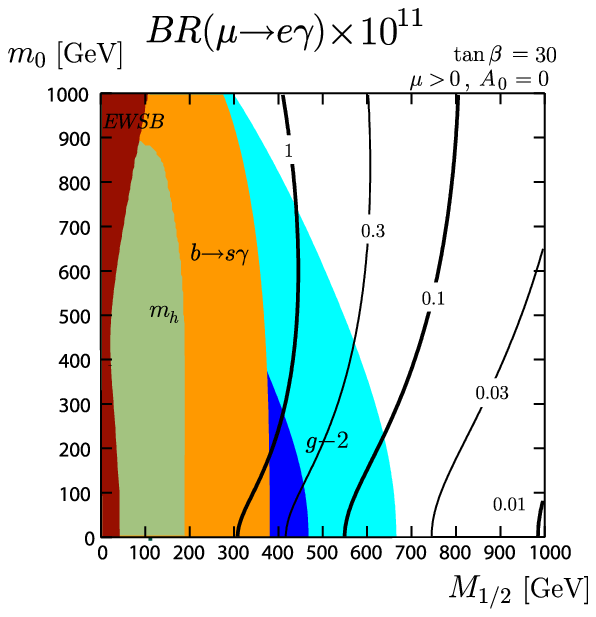}
\caption{Upper plot: contour plot in the $(m_0,M_{1/2})$ plane showing the values attained
by ${\rm BR}(\mu\to e\gamma)$ in the $SU(5)_{RN}$ model for $A_0=0$, $\tan\beta=10$ and $\mu>0$.
For the neutrino sector, we assume a hierarchical spectrum for both light and heavy neutrinos
and we take $m_{\nu_3}=0.05{\rm eV}$, $M_{3}=10^{13}\,{\rm GeV}$ and $U_{e3}=0.1$.
Lower plot: same as in the upper plot but for $\tan\beta=30$. In both plots, the grey region
is excluded by the constraint from the lower bound on the lightest Higgs boson mass $m_{h^0}$
(we impose $m_{h^0}>111.4$~GeV), the orange region is excluded by the constraints on
${\rm BR}(B\to X_s\gamma)$ at the $99\%$ C.L. limit, the light- blue (blue) region satisfies
$\Delta a^{\rm SUSY}_{\mu}> 1(2)\times 10^{-9}$ and finally the red region is excluded by the
requirement of a correct EWSB.}
\label{contour}
\end{figure}

In the upper (lower) plot of Fig.~\ref{contour}, we show the values reached by
${\rm BR}(\mu\to e\gamma)$ within the $SU(5)_{RN}$ model in the $(m_0,M_{1/2})$ plane setting
$\mu>0$, $A_0=0$ and $\tan\beta=10$ ($\tan\beta=30$). We assume $m_{\nu_3}=0.05{\rm eV}$, 
$M_{3}=10^{13}\,{\rm GeV}$ and $U_{e3}=0.1$. In both plots, the grey region is excluded
by the constraint from the lower bound on the lightest Higgs boson mass $m_{h^0}$ (we impose 
$m_{h^0}>111.4$~GeV), the orange region is excluded by the constraints on ${\rm BR}(B\to X_s\gamma)$
at the $99\%$ C.L. limit, the light blue (blue) region satisfies $\Delta a^{\rm SUSY}_{\mu}> 1(2)\times 10^{-9}$,
and finally the red region is excluded by the requirement of a correct electroweak symmetry breaking
(EWSB). We note that, passing from the case of $\tan\beta=10$ to the case of $\tan\beta=30$, the
indirect constraints, specially from $B\to X_s\gamma$, become stronger; however, the predictions
for both $\Delta a^{\rm SUSY}_{\mu}$ and ${\rm BR}(\mu\to e\gamma)$ increase while increasing
$\tan\beta$, so, as a final result, $\Delta a^{\rm SUSY}_{\mu}$ and ${\rm BR}(\mu\to e\gamma)$
reach large values even for heavy masses $(m_0,M_{1/2})\lesssim 1$~TeV. Moreover, we have found
that the requirement of a neutral lightest SUSY particle does not exclude any region in the 
$(m_0,M_{1/2})$ plane, in contrast to what happens in the constrained MSSM. The motivation is that,
within SUSY GUTs, the lightest stau is heavier than in the constrained MSSM because of GUT effects
stemming from the gauge interaction above the GUT scale, where the gauge couplings are unified.
%


Now let us comment about the prediction we would expect removing the assumptions $R=1$.
In the general case where $R\neq 1$, it turns out that $\delta^{\mu e}_{L}\sim\delta^{\tau \mu}_{L}$
and this leads to the following considerations: {\it i)} ${\rm BR}(\mu\to e\gamma)$ is always dominated 
by $\delta^{L}_{\mu e}$, hence we expect $A(\mu^+\to e^+\gamma)=+1$, {\it ii)} ${\rm BR}(\tau\to\mu\gamma)$ 
and $d_{\mu}$ are very suppressed because of the tight constraints from ${\rm BR}(\mu\to e\gamma)$.
Moreover, irrespective to whether $R=1$ or $R\neq 1$ and irrespective to the details of the light 
and heavy neutrino masses, the relation $\delta^{L}_{\mu e}\sim \delta^{L}_{\tau e}$ always holds 
thus implying a strong suppression for ${\rm BR}(\tau\to e\gamma)$ at the level of 
${\rm BR}(\tau\to e\gamma)/{\rm BR}(\mu\to e\gamma)\simeq 1/{\rm BR}(\tau\to e\nu_{\tau}\bar{\nu}_{e})$.

\begin{figure}
\includegraphics[scale=0.65]{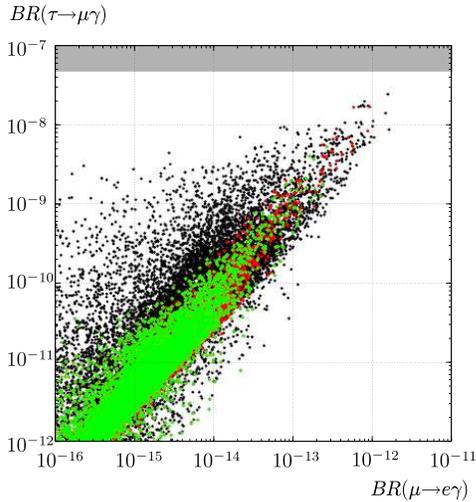}
\caption{
${\rm BR}(\mu\to e\gamma)$ vs ${\rm BR}(\tau\to \mu\gamma)$ in a pure SUSY $SU(5)$ model
without right-handed neutrinos. The plot has been obtained by means of a scan over the same
input parameters of Fig.~\ref{Fig:meg_dl}. Red and green dots satisfy the $B\to X_s\gamma$
constraints at the 99\% C.L. limit while black dots do not. Green dots additionally satisfy
$m_{h^0}>111.4$~GeV. All the points satisfy $\Delta a^{\rm SUSY}_{\mu}\leq 5\times 10^{-9}$.
The grey region is excluded by the current experimental upper bound on ${\rm BR}(\tau\to \mu\gamma)$.}
\label{Fig:meg_tmg2}
\end{figure}

Finally, we show ${\rm BR}(\mu\to e\gamma)$ vs ${\rm BR}(\tau\to \mu\gamma)$ in a pure SUSY $SU(5)$
model without right-handed neutrinos in Fig.~\ref{Fig:meg_tmg2}. As anticipated in previous sections,
these processes are typically quite suppressed as ${\rm BR}(\mu\to e\gamma)\lsim 10^{-(13-14)}$ and
${\rm BR}(\tau\to \mu\gamma)\lsim 10^{-(9-10)}$. However, there is a on-negligible fraction of points
lying within the experimentally interesting region where ${\rm BR}(\mu\to e\gamma)\gtrsim 10^{-13}$
and/or ${\rm BR}(\tau\to \mu\gamma)\gtrsim 10^{-9}$. This last situation happens only for large values
of $\tan\beta$ as ${\rm BR}(\ell_i\to \ell_j\gamma)\sim\tan^{2}\beta$.
Thus, a legitimate warning is whether in this region of the parameter space it is possible to satisfy
all the indirect constraints, specially those arising from processes enhanced by powers of $\tan\beta$
as $B\to X_s\gamma$, $B_s\to\mu^+\mu^-$, $B\to\tau\nu$, and the muon anomalous magnetic moment
$(g-2)$~\cite{IP}.
In the figure, the red and green dots satisfy the ${\rm BR}(B\to X_s\gamma)$ constraints at the
99\% C.L. limit.
As is well known, the charged Higgs contribution interferes constructively with the SM one while
the relative sign between the chargino and the SM amplitudes is given by $\hbox{sign}(A_t~\mu)$
as $A^{bsg}_{\tilde{\chi}^-}\propto [\mu A_t/m^{4}_{\tilde{q}}]\times \tan\beta$. Thus, to keep
$A_{\tilde{\chi}^-}$ under control with very large values of $\tan\beta$, we need large sfermion 
masses and small $A_t$.
Remembering that $A_t(m_t)\simeq 0.25 A_0 -2M_{1/2}$, we expect that small values for $A_t(m_t)$
are obtainable for positive and large $A_0$ compared to $M_{1/2}$ and this is exactly what we find 
numerically in the region where ${\rm BR}(\mu\to e\gamma)\gtrsim 10^{-13}$ and/or 
${\rm BR}(\tau\to \mu\gamma)\gtrsim 10^{-9}$. Moreover, the overall size for the total SUSY
amplitude is also reduced by means of cancellations between charged Higgs and chargino contributions.
Notice also that when $A_0$ is large, ${\rm BR}(\ell_i\to \ell_j\gamma)$ is enhanced as 
${\rm BR}(\ell_i\to \ell_j\gamma)\sim |\delta^{ij}|^2$ with the MI parameters $\delta^{ij}\sim (3m^2_0+A^2_0)$.

Concerning ${\rm BR}(B_s\to \mu^+\mu^-)$, we remind that its dominant amplitude is approximately given
by $A_{\tilde{\chi}^-}\propto [\mu A_t/m^{2}_{\tilde{q}}]\times [\tan^{3}\beta/M^{2}_{A}]$
hence, $A_{\tilde{\chi}^-}$ can be taken under control for small enough $A_t$ values and this is
already guaranteed by the constraints from ${\rm BR}(B\to X_s\gamma)$. 

In contrast to $B_s\to\mu^+\mu^-$ and $B\to X_s\gamma$, $\Btaun$ receives NP effects
already at the tree level by the charged Higgs exchange. These effects are particularly 
enhanced when $\tan\beta$ is large and if the heavy Higgs is light. However, we find that
$B\to\tau\nu$ receives sizable but small enough NP effects as for $\tan\beta\sim 40$ we 
find a quite heavy charged Higgs, i.e. $M_{H^+}\gtrsim 400~$GeV.

Likewise, it is easy to find the SUSY contributions to the muon anomalous magnetic moment of
the required size to explain its discrepancy with the SM expectation 
$\Delta a_{\mu}=a_{\mu}^{\rm exp}- a_{\mu}^{\rm SM}\approx(3 \pm 1)\times 10^{-9}$. This discrepancy
can be accommodated only with a positive $\mu$ sign, in agreement with the $B\to X_s\gamma$ requirements.

%

\section{Conclusions}

Motivated by the running MEG experiment, that will achieve an impressive resolution on the
branching ratio of $\mu\to e\gamma$ at the level of ${\rm BR}(\mu\to e\gamma)\lesssim 10^{-13}$,
in this article we have addressed the phenomenological implications, within supersymmetric
scenarios, of {\it i)} an observation of $\mu\to e\gamma$, {\it ii)} a significant improvement
of the ${\rm BR}(\mu\to e\gamma)$ upper bound.

In particular, we have exploited the correlations among ${\rm BR}(\mu\to e\gamma)$, the leptonic
electric dipole moments (EDMs) and the $(g-2)$ of the muon both in a model-independent way, {\it i.e.}
without making any assumption about the origin of the soft SUSY breaking terms, and in a specific but
more predictive scenario such as a supersymmetric $SU(5)$ model with right handed neutrinos.

In the following, we summarize the main results of our model-independent analysis:

\begin{itemize}
 \item The desire of an explanation for the muon $(g-2)$ anomaly
       $\Delta a_{\mu}\!=\!a_{\mu}^{\rm exp}\!-\!a_{\mu}^{\rm SM}\approx(3\pm 1)\times 10^{-9}$
       in terms of SUSY effects, leads to values for
       ${\rm BR}(\mu\to e\gamma)$ well within the MEG resolutions even for extremely tiny
       flavor mixing angles of order $\delta_{e\mu}\sim 10^{-5}$. This implies that the MEG
       sensitivities will enable us to test or to exclude a wide class of models predicting
       larger mixing angles.

 \item  Since the leptonic EDMs as induced by flavor effects are closely related to LFV processes
        as $l_i\to l_j\gamma$, an experimental evidence for $\mu\to e\gamma$ could likely imply
        large leptonic EDMs, well within their planned experimental resolutions (for the electron
        EDM, at least). In case both $\mu\to e\gamma$ and the electron EDM will be observed, their
        correlation will provide a precious tool to disentangle among the soft SUSY breaking terms
        violating the lepton flavor.

\end{itemize}

Concerning the analysis within a SUSY $SU(5)$ model with right handed neutrinos
we have found that

\begin{itemize}

 \item A SUSY contribution to the $(g-2)$ of the muon at the level of
       $\Delta a^{\rm SUSY}_{\mu}\approx(3\pm 1)\times 10^{-9}$ leads to values for
       ${\rm BR}(\mu\to e\gamma)$ well within the MEG sensitivities even when the
       unknown neutrino mixing angle $U_{e3}$ (to which ${\rm BR}(\mu\to e\gamma)$
       is very sensitive) is very small at the level of $U_{e3}<10^{-3}$.

\item  The predictions for the electron EDM typically lie above the value $d_e\gtrsim 10^{-30}e\,$cm
       for ${\rm BR}(\mu\to e\gamma)\gtrsim 10^{-13}$.

\item In case $\mu\to e\gamma$ would be observed, the knowledge of the P-odd asymmetry
      $A(\mu^{+}\to e^{+}\gamma)$ defined by means of initial muon polarization would represent
      a crucial tool to shed light on the nature of the LFV sources, in particular to disentangle
      whether an underlying SUSY GUT theory is at work or not. In fact, the pure MSSM with
      right-handed neutrinos unambiguously predicts that $A(\mu^{+}\to e^{+}\gamma)=+1$ while a SUSY
      $SU(5)$ model with right-handed neutrinos predicts $A(\mu^{+}\to e^{+}\gamma)\in (-1, +1)$.

\item An experimental evidence for $\mu\to e\gamma$ with a corresponding $A(\mu^+\to e^+\gamma)$ 
      departing sizably from $A(\mu^+\to e^+\gamma)=+1$ would most likely imply large (visible)
      values for ${\rm BR}(\tau\to \mu\gamma)$.

\item Both $\mu\to e\gamma$ and $\tau\to\mu\gamma$ turn out to be very sensitive probe of LFV
      effects arising in SUSY $SU(5)$ models with right handed neutrinos.
      While ${\rm BR}(\tau\to\mu\gamma)$ is not sensitive to $U_{e3}$, the predictions for
      ${\rm BR}(\mu\to e\gamma)$ are strongly affected by the unknown value of $U_{e3}$.
      As a result, both ${\rm BR}(\mu\to e\gamma)$ and ${\rm BR}(\tau\to\mu\gamma)$ can turn
      out to be the best probes of LFV in SUSY theories.

\end{itemize}

  In conclusion, the outstanding experimental sensitivities of the MEG experiment searching for
  $\mu\to e\gamma$, may provide a unique opportunity to get the first evidence of New Physics
  in low-energy flavor processes. Should this happen, we have outlined, within SUSY theories, 
  those low-energy observables that are also likely to show New Physics signals. Most importantly,
  a correlated study of the processes we have discussed in this work would represent a crucial
  step towards a deeper understanding of the underlying New Physics theory that is at work.

\textit{Acknowledgments:}

The work of JH is supported by the Grant-in-Aid for Scientific research from the Ministry of 
Education, Science, Sports, and Culture (MEXT), Japan, No.~20244037 and No.~2054252 (JH), and 
also by the World Premier International Research Center Initiative (WPI Initiative), MEXT, Japan.
The work of MN and PP was supported in part by the Cluster of Excellence ``Origin and Structure
of the Universe'' and by the German Bundesministerium f{\"u}r Bildung und Forschung under contract
05HT6WOA.


\begin{thebibliography}{999}

\bibitem{LFVnu}
T.~P.~Cheng and L.~F.~Li,
  Phys.\ Rev.\  D {\bf 16}, 1565 (1977);
S.~T.~Petcov,
  Sov.\ J.\ Nucl.\ Phys.\  {\bf 25}, 340 (1977)
  [Yad.\ Fiz.\  {\bf 25}, 641 (1977\ ERRAT,25,698.1977\ ERRAT,25,1336.1977)];
W.~J.~Marciano and A.~I.~Sanda,
  Phys.\ Lett.\  B {\bf 67}, 303 (1977);
B.~W.~Lee and R.~E.~Shrock,
  Phys.\ Rev.\  D {\bf 16}, 1444 (1977).

\bibitem{review}
  M.~Raidal {\it et al.},
  Eur.\ Phys.\ J.\  C {\bf 57}, 13 (2008)
  [arXiv:0801.1826 [hep-ph]].

\bibitem{meg}
Talk given by Marco Grassi, Les Rencontres de Physique de la Vallee D'Aoste,
La Thuile, Aosta Valley, Italy 
(March 1-7, 2009). Presentatation is placed on
http://agenda.infn.it/conferenceDisplay.py?confId=930.


\bibitem{bkl}
  K.~S.~Babu and C.~Kolda,
  Phys.\ Rev.\ Lett.\  {\bf 89} (2002) 241802
  [hep-ph/0206310].

\bibitem{paradisiH}
  P.~Paradisi,
  JHEP {\bf 0602}, 050 (2006) [arXiv:hep-ph/0508054];
  JHEP {\bf 0608}, 047 (2006) [arXiv:hep-ph/0601100];
  A.~Masiero, P.~Paradisi and R.~Petronzio,
  Phys.\ Rev.\  D {\bf 74}, 011701 (2006) [arXiv:hep-ph/0511289];
  A.~Dedes, J.~R.~Ellis and M.~Raidal,
  Phys.\ Lett.\  B {\bf 549} (2002) 159 [arXiv:hep-ph/0209207];
  A.~Brignole and A.~Rossi,
  Phys.\ Lett.\  B {\bf 566}, 217 (2003) [arXiv:hep-ph/0304081];
  Nucl.\ Phys.\  B {\bf 701} (2004) 3 [arXiv:hep-ph/0404211];
  R.~Kitano et al.,
  Phys.\ Lett.\  B {\bf 575} (2003) 300 [arXiv:hep-ph/0308021];
  E.~Arganda et al.,
  Phys.\ Rev.\  D {\bf 71}, 035011 (2005) [arXiv:hep-ph/0407302];
  JHEP {\bf 0806}, 079 (2008) [arXiv:0803.2039 [hep-ph]].


\bibitem{MFV}
  L.~J.~Hall and L.~Randall,
  Phys.\ Rev.\ Lett.\  {\bf 65}, 2939 (1990);
  G.~D'Ambrosio, G.~F.~Giudice, G.~Isidori and A.~Strumia,
  Nucl.\ Phys.\  B {\bf 645}, 155 (2002);
  P.~Paradisi, M.~Ratz, R.~Schieren and C.~Simonetto,
  Phys.\ Lett.\  B {\bf 668}, 202 (2008) [arXiv:0805.3989 [hep-ph]];
  G.~Colangelo, E.~Nikolidakis and C.~Smith,
  arXiv:0807.0801 [hep-ph].

\bibitem{Hall:1985dx}
  L.~J.~Hall, V.~A.~Kostelecky and S.~Raby,
  Nucl.\ Phys.\  B {\bf 267}, 415 (1986).

\bibitem{Borzumati:1986qx}
  F.~Borzumati and A.~Masiero,
   Phys.\ Rev.\ Lett.\  {\bf 57}, 961 (1986).

\bibitem{Hisano:1995cp}
  J.~Hisano, T.~Moroi, K.~Tobe, M.~Yamaguchi and T.~Yanagida,
  Phys.\ Lett.\  B {\bf 357}, 579 (1995)
  [arXiv:hep-ph/9501407];
  J.~Hisano, T.~Moroi, K.~Tobe and M.~Yamaguchi,
  Phys.\ Rev.\  D {\bf 53}, 2442 (1996)
  [arXiv:hep-ph/9510309].

\bibitem{Hisano:1998fj}
  J.~Hisano and D.~Nomura,
  Phys.\ Rev.\  D {\bf 59}, 116005 (1999)
  [arXiv:hep-ph/9810479].

\bibitem{Casas:2001sr}
  J.~A.~Casas and A.~Ibarra,
  Nucl.\ Phys.\  B {\bf 618}, 171 (2001)
  [arXiv:hep-ph/0103065].

\bibitem{calibbi}
  L.~Calibbi, A.~Faccia, A.~Masiero and S.~K.~Vempati,
  Phys.\ Rev.\  D {\bf 74} (2006) 116002 [arXiv:hep-ph/0605139].

\bibitem{Maltoni:2004ei}
M.~Maltoni, T.~Schwetz, M.~A.~Tortola and J.~W.~F.~Valle,
New J.\ Phys.\  {\bf 6} (2004) 122 [arXiv:hep-ph/0405172];
W.-M. Yao et al. (Particle Data Group), J. Phys. G 33, 1 (2006).

\bibitem{seesaw}
 S.~Davidson and A.~Ibarra,
  JHEP {\bf 0109}, 013 (2001)
  [arXiv:hep-ph/0104076];
 J.~R.~Ellis, J.~Hisano, M.~Raidal and Y.~Shimizu,
  Phys.\ Rev.\  D {\bf 66}, 115013 (2002)
  [arXiv:hep-ph/0206110].

\bibitem{Amsler:2008zzb}
  C.~Amsler {\it et al.}  [Particle Data Group],
  Phys.\ Lett.\  B {\bf 667}, 1 (2008).

\bibitem{Akeroyd:2004mj}
  A.~G.~Akeroyd {\it et al.}  [SuperKEKB Physics Working Group],
  arXiv:hep-ex/0406071;
  M.~Bona {\it et al.},
  arXiv:0709.0451 [hep-ex].

\bibitem{Barbieri:1994pv}
  R.~Barbieri and L.~J.~Hall,
  Phys.\ Lett.\ B {\bf 338}, 212  (1994) 
  [arXiv:hep-ph/9408406].

\bibitem{Barbieri:1995tw}
  R.~Barbieri, L.~J.~Hall and A.~Strumia,
  Nucl.\ Phys.\ B {\bf 445}, 219 (1995)
  [arXiv:hep-ph/9501334].

\bibitem{Hisano:1996qq}
  J.~Hisano, T.~Moroi, K.~Tobe and M.~Yamaguchi,
  Phys.\ Lett.\  B {\bf 391}, 341 (1997)
  [Erratum-ibid.\  B {\bf 397}, 357 (1997)] [arXiv:hep-ph/9605296].

\bibitem{Hisano:1997tc}
  J.~Hisano, D.~Nomura and T.~Yanagida,
  Phys.\ Lett.\  B {\bf 437}, 351 (1998) [arXiv:hep-ph/9711348];
  P.~Paradisi,
  JHEP {\bf 0510} (2005) 006 [arXiv:hep-ph/0505046].

\bibitem{Georgi:1979df}
  H.~Georgi and C.~Jarlskog,
  Phys.\ Lett.\  B {\bf 86}, 297 (1979).

\bibitem{Sakai:1981pk}
  N.~Sakai and T.~Yanagida,
  Nucl.\ Phys.\  B {\bf 197}, 533 (1982);
S.~Weinberg,
  Phys.\ Rev.\  D {\bf 26}, 287 (1982);
  P.~Nath, A.~H.~Chamseddine and R.~L.~Arnowitt,
  Phys.\ Rev.\  D {\bf 32}, 2348 (1985);
R.~L.~Arnowitt, A.~H.~Chamseddine and P.~Nath,
  Phys.\ Lett.\  B {\bf 156} 215 (1985);
 J.~Hisano, H.~Murayama and T.~Yanagida,
  Nucl.\ Phys.\  B {\bf 402}, 46 (1993)
  [arXiv:hep-ph/9207279].

\bibitem{Goto:1998qg}
  T.~Goto and T.~Nihei,
  Phys.\ Rev.\  D {\bf 59}, 115009 (1999)
  [arXiv:hep-ph/9808255];
 H.~Murayama and A.~Pierce,
  Phys.\ Rev.\  D {\bf 65}, 055009 (2002)
  [arXiv:hep-ph/0108104].

\bibitem{pqextend}
N.~Sakai and T.~Yanagida in Ref.~[15];
J.~Hisano, T.~Moroi, K.~Tobe and T.~Yanagida,
  Phys.\ Lett.\  B {\bf 342}, 138 (1995)
  [arXiv:hep-ph/9406417].

\bibitem{Hall:2001pg}
  L.~J.~Hall and Y.~Nomura,
  Phys.\ Rev.\  D {\bf 64}, 055003 (2001)
  [arXiv:hep-ph/0103125].

\bibitem{Bajc:2002bv}
  B.~Bajc, P.~Fileviez Perez and G.~Senjanovic,
  Phys.\ Rev.\  D {\bf 66}, 075005 (2002)
  [arXiv:hep-ph/0204311];
  D.~Emmanuel-Costa and S.~Wiesenfeldt,
  Nucl.\ Phys.\  B {\bf 661}, 62 (2003)
  [arXiv:hep-ph/0302272].

\bibitem{ArkaniHamed:1995fs}
  N.~Arkani-Hamed, H.~C.~Cheng and L.~J.~Hall,
  Phys.\ Rev.\  D {\bf 53}, 413 (1996)
  [arXiv:hep-ph/9508288];
  J.~Hisano, D.~Nomura, Y.~Okada, Y.~Shimizu and M.~Tanaka,
  Phys.\ Rev.\  D {\bf 58}, 116010 (1998)
  [arXiv:hep-ph/9805367].


\bibitem{Hisa1}
  J.~Hisano and K.~Tobe,
  Phys.\ Lett.\ B {\bf 510}, 197 (2001) 
  [hep-ph/0102315].

\bibitem{g_2_th}
  M.~Passera,
  J.\ Phys.\ G {\bf 31} (2005) R75
  [hep-ph/0411168];
  Nucl.\ Phys.\ Proc.\ Suppl.\  {\bf 155} (2006) 365
  [hep-ph/0509372];
  M.~Davier,
  Nucl.\ Phys.\ Proc.\ Suppl.\  {\bf 169}, 288 (2007)
  [arXiv:hep-ph/0701163];
  K.~Hagiwara, A.~D.~Martin, D.~Nomura and T.~Teubner,
  Phys.\ Lett.\  B {\bf 649}, 173 (2007)
  [arXiv:hep-ph/0611102].

\bibitem{g_2_exp} 
   H.~N.~Brown {\it et al.}  [Muon g-2 Collaboration],
  Phys.\ Rev.\ D {\bf 62},  091101 (2000)
  [hep-ex/0009029];
   Phys.\ Rev.\ Lett.\ {86}{2001}{2227} [hep-ex/0102017];
  Phys.\ Rev.\ Lett.\  {\bf 89},  101804 (2002)
  [hep-ex/0208001];
  Phys.\ Rev.\ Lett.\  {\bf 92}, 161802 (2004)
  [hep-ex/0401008].

\bibitem{passera_mh}
  M.~Passera, W.~J.~Marciano and A.~Sirlin,
  Phys.\ Rev.\  D {\bf 78}, 013009 (2008)
  [arXiv:0804.1142 [hep-ph]].



\bibitem{pospelov}
For a review of EDMs please see, M.~Pospelov and A.~Ritz,
Annals Phys.\  {\bf 318}, 119 (2005);
J.~R.~Ellis, J.~S.~Lee and A.~Pilaftsis,
JHEP {\bf 0810}, 049 (2008) [arXiv:0808.1819 [hep-ph]] 
and therein references.

\bibitem{HNP_EDM}
  J.~Hisano, M.~Nagai and P.~Paradisi,
  Phys.\ Lett.\  B {\bf 642}, 510 (2006)
  [arXiv:hep-ph/0606322];
  Phys.\ Rev.\  D {\bf 78}, 075019 (2008) [arXiv:0712.1285 [hep-ph]];
  arXiv:0812.4283 [hep-ph].



\bibitem{ABP}
W.~Altmannshofer, A.~J.~Buras and P.~Paradisi,
  Phys.\ Lett.\  B {\bf 669}, 239 (2008)
  [arXiv:0808.0707 [hep-ph]].

\bibitem{edminseesaw}
  J.~R.~Ellis, J.~Hisano, M.~Raidal and Y.~Shimizu,
  Phys.\ Lett.\  B {\bf 528}, 86 (2002)
  [arXiv:hep-ph/0111324];
  I.~Masina,
  Nucl.\ Phys.\  B {\bf 671}, 432 (2003)
  [arXiv:hep-ph/0304299];
 Y.~Farzan and M.~E.~Peskin,
  Phys.\ Rev.\  D {\bf 70}, 095001 (2004)
  [arXiv:hep-ph/0405214].

\bibitem{96KunoOkada}
 Y.~Kuno and Y.~Okada, Phys.~Rev.~Lett.~{\bf 77}, {3} (1996);
 Y.~Kuno, A~.Maki and Y.~Okada, Phys.~Rev.~{\bf D}{\bf 55}, {2517} (1997).

\bibitem{Okada:1999zk}
  Y.~Okada, K.~i.~Okumura and Y.~Shimizu,
  Phys.\ Rev.\  D {\bf 61}, 094001 (2000)
  [arXiv:hep-ph/9906446].

\bibitem{gut_FCNC}
  T.~Moroi,
  Phys.\ Lett.\  B {\bf 493}, 366 (2000);
  J.~Hisano and Y.~Shimizu,
  Phys.\ Lett.\  B {\bf 565}, 183 (2003);
  J.~Hisano and Y.~Shimizu,
  Phys.\ Lett.\  B {\bf 581}, 224 (2004)
  [arXiv:hep-ph/0308255];
  M.~Ciuchini et al.,
  Phys.\ Rev.\ Lett.\  {\bf 92}, 071801 (2004);
  M.~Ciuchini et al.,
  Nucl.\ Phys.\  B {\bf 783}, 112 (2007);
  T.~Goto, Y.~Okada, T.~Shindou and M.~Tanaka,
  Phys.\ Rev.\  D {\bf 77}, 095010 (2008);
  J.~Hisano and Y.~Shimizu,
  Phys.\ Lett.\  B {\bf 669}, 301 (2008);
  P.~Ko, J.~h.~Park and M.~Yamaguchi,
  arXiv:0809.2784 [hep-ph].

\bibitem{gutedm}
S.~Dimopoulos and L.~J.~Hall,
  Phys.\ Lett.\ B {\bf 344}, 185  (1995);
R.~Barbieri, A.~Romanino and A.~Strumia,
  Phys.\ Lett.\ B {\bf 369}, 283  (1996);
A.~Romanino and A.~Strumia,
  Nucl.\ Phys.\ B {\bf 490}, 3  (1997);
J.~Hisano, M.~Kakizaki, M.~Nagai and Y.~Shimizu,
  Phys.\ Lett.\ B {\bf 604}, 216 (2004);
J.~Hisano, M.~Kakizaki and M.~Nagai,
  Phys.\ Lett.\ B {\bf 624}, 239 (2005).


\bibitem{mu2e} R.~Bernstein, talk given in the 4th International Workshop
  on Nuclear and Particle Physics at J-PARC (NP08), Mito, Ibaraki,
  Japan, March, 2008 (http://j-parc.jp/NP08/).

\bibitem{comet} A.~Sato, talk given in the 4th International Workshop
  on Nuclear and Particle Physics at J-PARC (NP08), Mito, Ibaraki,
  Japan, March, 2008 (http://j-parc.jp/NP08/).

\bibitem{IP}
  For an extensive analysis of these observables please see, G.~Isidori and P.~Paradisi,
  Phys.\ Lett.\ B {\bf 639},  499 (2006);
  G.~Isidori, F.~Mescia, P.~Paradisi and D.~Temes,
  Phys.\ Rev.\  D {\bf 75} (2007) 115019;
  G.~Barenboim, P.~Paradisi, O.~Vives, E.~Lunghi and W.~Porod,
  JHEP {\bf 0804}, 079 (2008).

\bibitem{misiak}
  M.~Misiak {\it et al.},
  Phys.\ Rev.\ Lett.\  {\bf 98}, 022002 (2007);
  M.~Misiak and M.~Steinhauser,
  Nucl.\ Phys.\  B {\bf 764}, 62 (2007).

\end{thebibliography}
\end{document}